\documentclass[aps,pre,amsmath,amssymb,reprint]{revtex4-1}
\usepackage{CJK}
\usepackage{amssymb,euscript,latexsym,amsmath}
\usepackage{graphicx}
\usepackage[dvips]{epsfig}
\usepackage{color}
\usepackage{epstopdf}
\usepackage{setspace}
\usepackage{subfigure}
\usepackage{chngpage}
\usepackage{hyperref}

\usepackage[T1]{fontenc}
\usepackage[utf8]{inputenc}
\usepackage[font=small,labelfont=bf,tableposition=top]{caption}
\usepackage{booktabs,multirow}

\DeclareGraphicsExtensions{.eps,.ps,.pdf}

\begin{document}

\title{Design Principles for Biological and Mathematical Cilia}
\title{Design Principles for Robust Ciliary Peformance}
\title{Design Principles for the Best Performing Cilium: Robustness versus Optimality}
%\title{Design Principles for Robust Ciliary Peformance}
\title{The Best Performing Cilium:  Efficient or Robust?}
\title{The Most Efficient Cilium is not Most Robust to Perturbations}
\title{Evaluating efficiency and robustness in cilia design}

\author{Hanliang Guo}
\author {Eva Kanso}
\email{kanso@usc.edu}
\affiliation{Department of Aerospace and Mechanical Engineering,  \\ University of Southern California, Los Angeles, California 90089, USA}
%\affil{Department of Aerospace and Mechanical Engineering, \\University of Southern California\\Los Angeles, California, USA}

\date{December 20, 2015}

\begin{abstract}
Motile cilia are used by many eukaryotic cells to transport flow. 
Cilia-driven flows are important to many physiological functions, yet a deep understanding of the interplay between the mechanical structure of cilia and their physiological functions in healthy and diseased conditions remains elusive. For developing such understanding, one needs a quantitative framework for assessing  cilia performance and robustness when subject to perturbations in the cilia apparatus. Here, we link cilia design (beating patterns) to function (flow transport) in the context of experimentally- and theoretically-derived cilia models. We particularly examine the optimality and robustness of cilia design. Optimality refers to efficiency of flow transport, while robustness is defined as low sensitivity to variations in the design parameters. We find that suboptimal designs can be more robust than optimal ones. That is, designing for
the most efficient cilium does not guarantee robustness. These findings have significant implications on the understanding of cilia design in artificial and biological systems.

%Our work provides tools for investigating cilia performance under various operating conditions and, thus, have significant implications on the understanding of cilia design in artificial and biological systems.
%In a variety of biological processes, eukaryotic cells use cilia to transport flow. 
%Although cilia have a
%remarkably conserved internal molecular structure, experimental observations report very diverse kinematics.
%To address this diversity, we determine numerically the kinematics and energetics of the most
%efficient cilium. Specifically, we compute the time-periodic deformation of a wall-bound elastic filament
%leading to transport of a surrounding fluid at minimum energetic cost, where the cost is taken to be the
%positive work done by all internal molecular motors. The optimal kinematics are found to strongly depend
%on the cilium bending rigidity through a single dimensionless number, the Sperm number, and closely
%resemble the two-stroke ciliary beating pattern observed experimentally. 
%
\end{abstract}

%the notion that hydrodynamic effciency dictates the cilia beating kinematics,
%and suggest that other biological functions and constraints play a role in explaining
%the wide variety of cilia beating patterns observed in biological systems.
%
%the notion that hydrodynamic eciency dictates the cilia beating kinematics,
%and suggest that other biological functions and constraints play a role in explaining
%the wide variety of cilia beating patterns observed in biological systems.
\maketitle

\singlespacing

\section{Introduction}
Cilia are hair-like structures, {typically tens of} micrometers in length, that protrude from the surface of eukaryotic cells. They can be found in many aquatic and terrestrial species from the unicellular protozoan \textit{Paramecium} to humans. 
%Cilia could occur along both external and internal epithelial surfaces in aquatic species  but  are restricted to internal epithelial surfaces in terrestrial species. 
In mammals, cilia are present on many cells in the body, either in large groups on a single cell, as in the case of motile cilia, or as solitary structures, as in the case of primary and nodal cilia \cite{Christensen2007}.
Motile cilia, the focus of this work, are found on the epithelial cells of the trachea~\cite{fulford1986, ocallaghan_respiratory_1999,randell_effective_2006}, ependymal cells in the brain~\cite{DelBigio1995,Mirzadeh2010}, and  cells lining the oviduct and epididymis of the reproductive tracts~\cite{Lyons2006}.  They normally beat in an orchestrated fashion resulting in fluid movement and cell transport~\cite{wong_nature_1993}.
Great advances have been made  in demonstrating the importance of ciliary transport to many physiological functions~\cite{wanner_mucociliary_1996, davenport_incredible_2005} and in unraveling the underlying fluid-structure interactions at the cilia scale~\cite{chopra_measurement_1977,smith_modelling_2008, smith_mathematical_2009,li_methods_2012,ding2014}.   However, a deep understanding of the interplay between the   mechanical structure of cilia and their physiological functions and how ciliary dysfunction can lead to severe disease and developmental pathologies remains elusive. 

It is therefore important to apply quantitative measures  that link cilia mechanics (e.g., cilia beating patterns) to function (e.g., flow transport) in both healthy and diseased states~\cite{fliegauf2007, afzelius2004cilia}. The lack of such standardized measures is in part due to the use of disparate research approaches in the biological and physical sciences.  In health-related research, a traditional approach is to use {\em in vitro} ciliated cell cultures, from which the fundamental structure-function relationships are inferred. However, the clinical use of such  {\em in vitro} cell cultures has been mostly qualitative.  In biofluid mechanics research, an increasingly popular approach consists of computing the ideal kinematics for a single function, such as optimal fluid transport in cilia acting both individually~\cite{eloy2012} and collectively~\cite{osterman2011}. This optimization approach requires a sophisticated mathematical and computational apparatus to arrive at the optimal cilia kinematics. The optimal result provides valuable insights into ciliary design, but fails to explain, let alone evaluate, the variation in cilia kinematics and how deviations from such kinematics affect cilia function. {Here, we posit that robustness under perturbations to cilia kinematics, whether due to environmental or structural causes, is of paramount importance for healthy cilia function. As such, lack of robustness can be linked to cilia dysfunction and disease.} 
We therefore present an alternative approach  that emphasizes
%, rather than optimality itself, 
the design principles amenable to a given function (flow transport), and 
%how this transport function is affected 
the robustness of such function under variations in the cilia design parameters. 
%\textcolor{black}{examines the robustness of cilia function to cilia kinematics.}
%\textcolor{black}{Our main goal is to examine the sensitivity of cilia function, i.e., flow transport, to perturbations in the cilia beating kinematics.}
%

The beating cycle of a cilium typically consists of two phases: an effective stroke aimed at generating flow followed by a recovery stroke during which the cilium returns to its initial position. During the effective stroke, the cilium moves in an almost straight configuration in a plane normal to the cell surface, while in the recovery stroke, it bends parallel to the cell surface while exhibiting  large curvatures and possibly moving out of the normal plane. The details of the cilia beat kinematics depend on the cell type, but the exact physical and/or biological mechanisms that select or constrain these kinematics are not well understood. We posit that optimality is not the main mechanism driving this selection. 
This conviction is in part based on a recent comparative study of the performance of two cilia beating patterns taken from two experimental systems, \textcolor{black}{namely, cilia from a swimming microorganism and  rabbit tracheal cilia. In the first system, cilia are used for swimming while in the second, they are used for fluid transport.} \textcolor{black}{If cilia performance were
hydrodynamically optimal, one would expect the transport-specific cilia to outperform the swimming-specific
cilia in fluid transport, and vice versa.
However,} by comparing the two types of cilia, we found the cilia beating pattern taken from the swimming system to be consistently superior to the other in three different {hydrodynamic} performance metrics. 
%Specifically, the beating pattern taken from swimming system performs better than the one taken from fluid transport system, even in the simulated fluid transport system.} % \textcolor{black}{including the average swimming speed/transport flow rate, the maximum internal moments generated by the cilia, and swimming/transport efficiencies}.  
 These findings imply that cilia beating kinematics need not be optimal {hydrodynamically}~\cite{guo2014}. Further, in mammalian cells, cilia beating motion takes a minimal fraction of the metabolic budget of the body. It is therefore unlikely, from the perspective of evolutionary biology, that the energetic cost associated with the beating of a cilium has posed a major selective pressure on its beat kinematics. In contrast, the energetic cost associated with locomotion consumes a significant  portion of the metabolic budget in almost all motile organisms~\cite{tucker1970energetic}, thus justifying the need for an optimal or quasi-optimal gait. This is also true in cilia-driven locomotion such as in the protozoan \textit{Paramecium}~\cite{katsu2009}. The \textit{Paramecium} uses more than half its total energy consumption for swimming, while its {hydrodynamic} swimming efficiency is estimated to be as low as 0.77\%. But it is important to distinguish between optimizing the cilium kinematics and optimizing the swimming gait. The latter involves, in addition to the cilium kinematics, the coordination between multiple cilia and different cilia types as well as their distribution on the underlying surface and the geometric properties of that surface. 
 % \textcolor{black}{as well as the relative orientations of the flagella with respect to the cells in mono- and bi-flagellates, such as spermatozoon or \textit{Chlamydomonas}.} 

In this study, we examine the beating kinematics of individual cilia in relation to their  function in fluid transport.
We propose three reduced design parameters that capture the salient kinematic features of motile cilia, namely, the leaning angle in the direction of the effective stroke, the beating amplitude of the effective stroke, and the out-of-plane angle. We present a straightforward approach for extracting these design parameters from any cilia beating pattern, including those obtained experimentally from high-speed image sequences. We then present a mathematical family of cilia-like kinematics which can be constructed for any combination of design parameters. 

We validate this choice of ``generic'' cilia-like kinematics by comparing it to experimentally-derived cilia kinematics before we use it to investigate questions of optimality and robustness of cilia design. Optimality here is defined in terms of the efficiency of flow transport, while robustness refers to lack of sensitivity {in transport efficiency} to variations in the design parameters. We find that suboptimal kinematics are more robust than optimal kinematics. That is, there is an interplay between optimality and robustness in cilia design. To our knowledge, our study is the first to propose robustness as an important criterion in cilia design. Our findings  provide a quantitative framework for investigating cilia performance under various operating conditions and, thus, have significant implications on the understanding of cilia design in artificial and biological systems.

%--------------------
\begin{figure*}[t]
\begin{center}
{\includegraphics{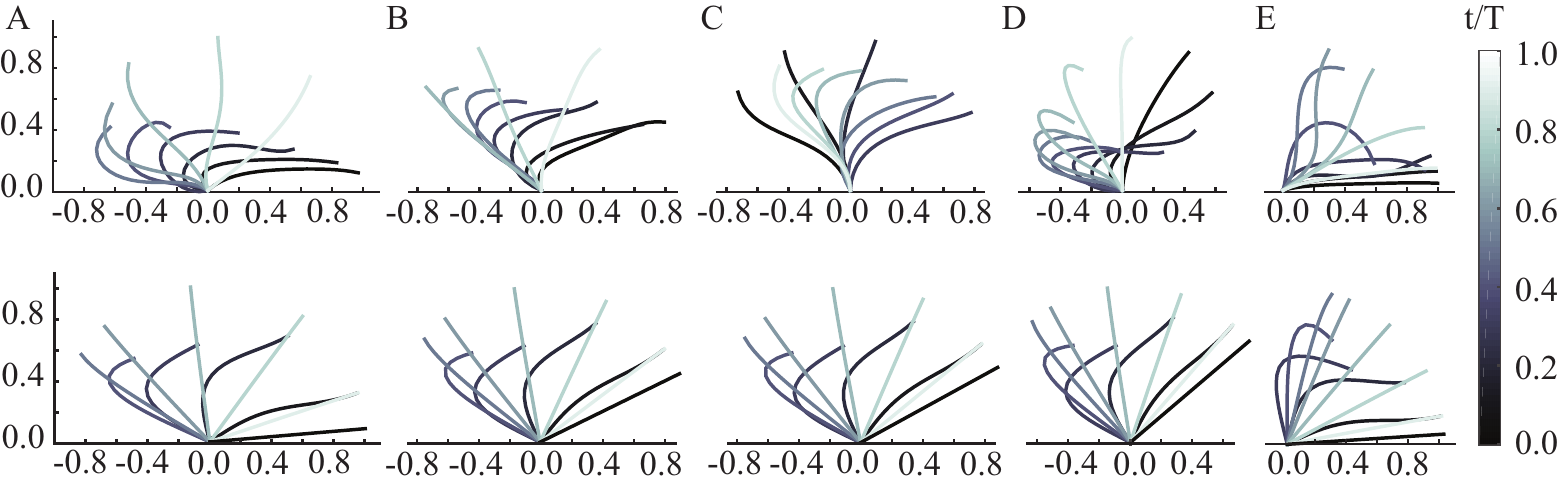}}
%{\includegraphics{figs/StrokePanelHSV.eps}}
\end{center}
  	\caption[]{Cilia beating patterns reconstructed from: (top row) experimental data  extracted from Sleigh~\cite{sleigh1968}, and Fulford \& Blake~\cite{fulford1986}. A: \textit{Didinium}; B: \textit{Paramecium}; C: Rabbit tracheal cilia; D: \textit{Sabellaria} gill; E: \textit{Opalina}; (bottom row) mathematical model presented in this paper where the values of $(\lambda_o, \alpha_o)$ are extracted from the experimental images in the top row.}
	\label{fig:strokes}
\end{figure*}

\section{Model and Method}

We consider a cilium whose base point is chosen to coincide with the origin of  properly-chosen Cartesian coordinates $(x, y, z)$. The length of the cilium is $l$.  The cilium beats periodically in the half infinite domain $y>0$ with frequency $\omega$ and period $T = 2\pi/\omega$. 

The  beating motion of the cilium is represented by $\mathbf{x}_c (s, t) \equiv  (x_c(s,t), y_c(s,t), z_c(s, t))$, where $s$ is the arclength along the cilium's centerline from its base $(0<s<l)$ and $t$ is time $(0<t<T)$. To describe the cilia beating kinematics from experimental observation of cilia beat patterns~\cite{sleigh1968,fulford1986}, we write each component of  $\mathbf{x}_c(s,t)$ using a Taylor series expansions in $s$ and Fourier series expansion in $t$. We then calculate the series coefficients that best fit the  experimental data~\cite{sleigh1968,fulford1986} subject to the constraint that the total length of the cilium is conserved,  see Fig.~\ref{fig:strokes}. 
Here,  for consistency and without loss of generality, we consider the effective stroke direction to be pointing in the positive $x$-direction. It is clear from Fig.~\ref{fig:strokes} that the details of the cilia beating kinematics vary depending on the cell type, but qualitatively they all follow the same trend. 
%------------------------
\begin{figure*}[t]
\begin{center}
{\includegraphics{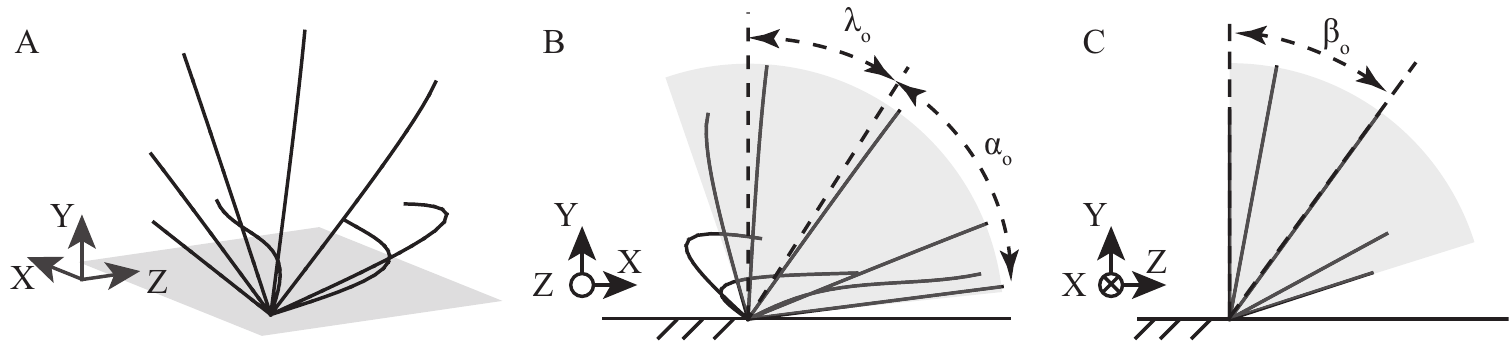}}
\end{center}
  	\caption[]{Reduced cilia design parameters $\lambda_o, \alpha_o$ and $\beta_o$. A:  three-dimensional beating pattern of the cilium. B:  projection onto the plane of the effective stroke showing the leaning angle $\lambda_o$ and beating amplitude $\alpha_o$. C: projection onto the plane normal to the effective stroke showing the out-of-plane swinging angle $\beta_o$.}
	\label{fig:abmodel}
\end{figure*}
%------------------------
Further, all these examples   correspond to planar cilia kinematics. 
Based on the original planar images in Sleigh~\cite{sleigh1968}, and Fulford \& Blake~\cite{fulford1986},  it is relatively straightforward to assess whether the bending kinematics is planar
%For each of these cilia, we quantified the out-of-plane component of the cilium's motion 
by comparing the projected cilium length in the experimental image sequence of the beat cycle to absolute cilium length~\cite{sleigh1968}. In each of the cases presented here, we found little change in the cilium's overall length over one beating cycle (with standard deviation about 3\% ), indicating that the cilium bending motion is mostly two-dimensional. Note that in the cases when the cilia kinematics is non-planar, it is very difficult to uniquely reconstruct the out-of-plane motion from planar images. For this reason, we only reconstructed planar kinematics in Fig.~\ref{fig:strokes}.

%Here, $\phi$ is the phase within each beating cycle. \ek{is there a particular reason to use $\phi$ and not $t$?} (In this paper, $\phi$ and $t$ are identical. However there could be a (non-linear) mapping between $\phi$ and $t$ to achieve non-equal effective/recovery strokes... )
%Note that it is relatively straightforward,  based on planar images, to assess whether the beating kinematics is planar, but it is very difficult to uniquely reconstruct the out-of-plane component when the beating kinematics is non-planar.  Therefore, we only reconstruct planar kinematics in~\ref{fig:strokes}.

%Here, we model the cilium as an inextensible elastic filament of length $l$ with one end attached to the origin of a Cartesian frame $(x,y,z)$.  The cilium beats periodically in the half infinite domain $y>0$. The beating time period is $T$. Without loss of generality, we consider the effective stroke direction to be pointing in the positive $x$-direction. Motion in the $z$-direction occurs when the beating pattern is  three-dimensional, see Fig.~\ref{fig:abmodel}.

In order to facilitate the comparison between the beating kinematics of various cilia type,
{it is useful to describe the cilia beating patterns mathematically using a small number of parameters}. {Concise mathematical descriptions date back to  G. I. Taylor's swimming sheet~\cite{taylor1951}, albeit for modeling symmetric beating motion.  Asymmetric beating patterns were later described using various approaches, including the ``biased baseline'' mechanisms \cite{eshel1987new}, the Taylor and Fourier series expansions~\cite{fulford1986} employed above, and more recently,  efficient polynomial expansions~\cite{bayly2010efficient} and  curvature-based Fourier series expansions~\cite{sartori2015dynamic}. These methods can reconstruct the beating patterns from given experimental data, but often require sophisticated machinery and large number of parameters. Here,}
we  introduce  three reduced ``cilia design'' parameters {that can be used to reconstruct any beating patterns in 3D}: the leaning angle $\lambda_o$, the beating amplitude $\alpha_o$, and the swinging angle $\beta_o$, as shown in Fig.~\ref{fig:abmodel}. To define these parameters, %Note that the swinging angle is $0$ for 2D beating patterns.
%Here $\lambda_o$ is a measure of how much the cilium leans in the fluid in average within a cycle; $\alpha_o$ measures the angular beating amplitude; $\beta_o$ measures the  
we first let $\lambda$ be  the angle between the straight line pointing from the cilium base to the cilium tip and the $(y,z)$ plane and $\beta$ be the angle between {the plane containing the curved cilium} and the $(x,y)$ plane. {Positive $\lambda$ are taken to be counter-clockwise about the positive $z$-axis, whereas positive $\beta$ are measured clockwise about the positive $x$-axis.} Basically, $\lambda$ describes the amount by which the cilium is leaning in the direction of the effective stroke, while $\beta$ describes the amount by which the cilium leans {away the plane of the effective stroke in the positive $z$-axis}. The three design parameters $\lambda_o$, $\alpha_o$ and $\beta_o$ can then be defined as follows:
%----
\begin{equation}
\label{eq:parameters}
\begin{split}
%\lambda_o = \frac{1}{2}[\max_{0\le\phi<T}(\lambda)+\min_{0\le\phi<T}(\lambda)],\\
%\alpha_o = \frac{1}{2}[\max_{0\le\phi<T}(\lambda)-\min_{0\le\phi<T}(\lambda)],\\
%\beta_o = \frac{1}{2}[\max_{0\le\phi<T}(\beta)+\min_{0\le\phi<T}(\beta)].\\
\lambda_o = \frac{1}{2}[\max_{0\le t<T}(\lambda)+\min_{0\le t <T}(\lambda)], \\
\alpha_o = \frac{1}{2}[\max_{0\le t<T}(\lambda)-\min_{0\le t<T}(\lambda)],\\
\beta_o = \frac{1}{2}[\max_{0\le t <T}(\beta)+\min_{0\le t <T}(\beta)].
\end{split}
\end{equation}
%----
For the examples reported in Fig.~\ref{fig:strokes}, $\beta$ and  $\beta_o$ are identically zero. {These three parameters capture the salient features of the actual beating patterns but not all their details. The main reason we choose these parameters is that they are easy to access from experimental images and they are amenable to a low-order mathematical representation of asymmetric cilia beating patterns as discussed next.}

%------------------------
\begin{figure*}[t]
\begin{center}
{\includegraphics{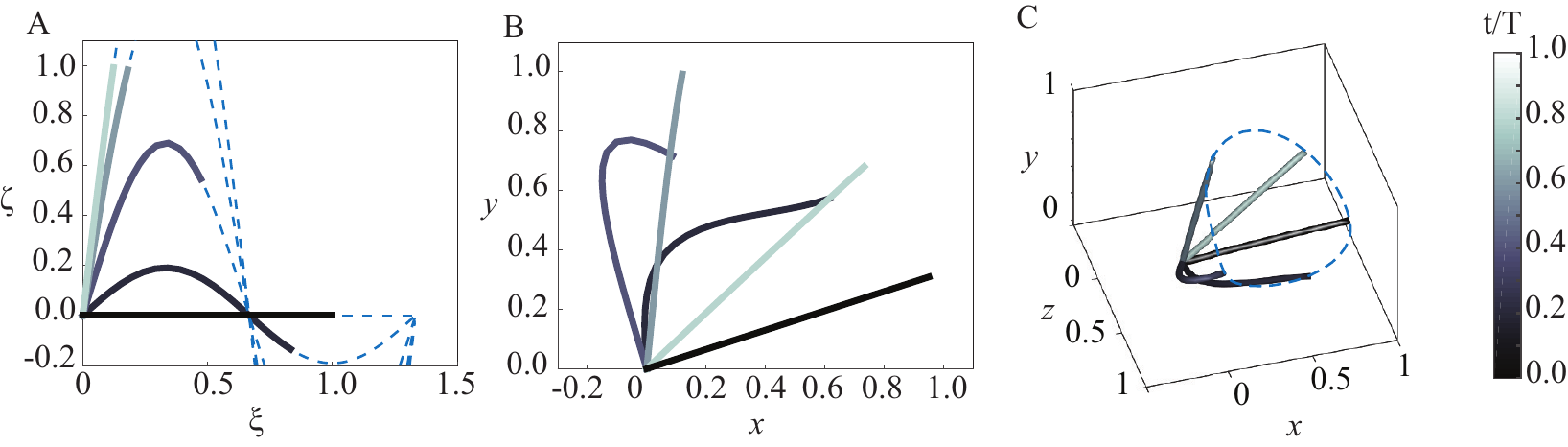}}
%{\includegraphics{figs/ModelPanelHSVV2.eps}}
\end{center}
  	\caption[]{Generic cilia-like kinematics. A: shapes of the strokes at select times are constructed from truncated sinusoidal functions that ensure conservation of total length of the cilium. B: planar beating patterns at select times. C: beating patterns at selected phases in three-dimensional space. Parameter values in all subfigures are  $\lambda_o = .2\pi$, $\alpha_o = .2\pi$, $\beta_o = .2\pi$}
	\label{fig:model}
\end{figure*}
%------------------------

We now introduce a mathematical family of functions that emulates cilia beating kinematics. This mathematical family can be viewed as a ``blue print'' for generating cilia beating kinematics that satisfy given cilia design parameters $(\lambda_o, \alpha_o, \beta_o)$. {Basically, we use a sinusoisal function to describe the shape of the cilium at different times in local Cartesian coordinates; we then rotate the shape according to the design parameters $\lambda_o$, $\alpha_o$ and $\beta_o$ to generate cilia-like kinematics.}
%Inspired by the real beating patterns shown in Fig.~\ref{fig:strokes}, we propose a family of generic beating strokes whose shapes are parts of a sinusoidal wave described by the following equation
%The generating function is a time-dependent sinusoidal wave of the form
%\textcolor{black}{The shape of the cilium is defined as a time-dependent sinusoidal wave of the form}
To this end, we let
%----
\begin{equation}	
\label{eq:stroke}
\zeta(\xi,t) = lA(t) \sin\left(\frac{3\pi}{2}\frac{\xi}{l}\right), \quad  A(t) = 1 -\cos\left(\pi\frac{t}{T}\right).
\end{equation}
%----
%\eknote{I assume you mean $y(x) = Al \sin(\frac{3\pi}{2}\frac{x}{l})$?}
Here, $(\xi,\zeta)$ are {local Cartesian coordinates} and must satisfy the constraint that the total length of the cilium is constant for all time $t$. Namely, one has $\mathrm{d}s = \sqrt{(\mathrm{d}\xi)^2 + (\mathrm{d}\zeta)^2}$, which yields $\int_0^l \mathrm{d}s = \int_0^l  \mathrm{d}{\xi} \sqrt{1+ (\mathrm{d}\zeta/\mathrm{d}\xi)^2} = l $. By virtue of the implicit function theorem, one can use this equality together with \eqref{eq:stroke} to write $\xi(s,t)$ and $\zeta(s,t)$ as functions of arc-length  and time. Fig.~\ref{fig:model}A  is a depiction of the kinematics obtained from these  generating functions. %\eknote{include reference to figure 3A}
{The wave number $3\pi/2$  is  chosen so that the cilium  does not exhibit both positive and negative curvatures at any given time, consistent with experimental observation of cilia beating patterns (see Fig.~\ref{fig:strokes} and analysis in \cite{sleigh1968}).
The time-dependent wave amplitude $A(t)$ is chosen so that the cilium is less curved in the effective stroke ($0.5T<t<T$) than in the recovery stroke ($0<t<0.5T$). Note that one can expand this representation in terms of a 2$^{nd}$ order Taylor expansion in $s$, given that no inflection points in the cilium shape are allowed, and a Fourier series expansions in time.  The coefficients of these expansions are omitted here for brevity. 
%Inversely, starting from the Taylor and Fourier series expansions of~\cite{sleigh1968,fulford1986}, one can readily construct this low-order mathematical representation by first obtaining calculating the values $(\lambda_o, \alpha_o, \beta_o)$ and using the following procedure.
}
%%----
%\begin{equation}	\label{stroke}
%0\le \int_0^x \sqrt{1+ (dy/dx)^2}\, \mathrm{d}\tilde{x}\le l, \ek{\textrm{I have no idea what this means! it is not an equality constraint}}
%\end{equation}
%%----
% where the phase dependent wave amplitude $A(\phi) = 1-\cos(\frac{\pi}{T}\phi)$.  
%\ek{what do you mean by: The wave is truncated when the arc-length reaches 1 to preserve the cilium length? can you describe mathematically what you are trying to say} 
%The generic beating stroke shapes are shown in Fig.~\ref{fig:model}A. To get the actual beating patterns at phase $\phi$, rotate the stroke about the cilium base such that $\lambda(\phi) = \alpha_o\cos(\frac{2\pi}{T}\phi)+\lambda_o$ and $\beta(\phi) = \beta_o[\sin(\frac{2\pi}{T}\phi)+1]$. 

Cilia-like kinematics with desired values of $(\lambda_o, \alpha_o, \beta_o)$  can be constructed using the ``blue print'' $\left(\xi(s,t),\zeta(s,t)\right)$ as follows.
%\ek{this is not helping me much! You want to start from  $(\lambda_o, \alpha_o, \beta_o)$ and the ``blue print'' $y(x,t)$ but you want to obtain mathematical expression for $x$, $y$ and $z$} For any given design parameters $(\lambda_o, \alpha_o, \beta_o)$ and phase $\phi$, 
We first write the {instantanous} leaning angle $\lambda(t)$  and swinging angle $\beta(t)$ as explicit functions of time
%The time-dependent leaning angle is given we first rotate the stroke about the $z$-axis such that
%----
\begin{equation}
\begin{split}
\lambda(t) &= \lambda_o + \alpha_o\cos\left(2\pi\frac{t}{T}\right), \\
\beta(t) &= \beta_o\left[1 + \sin\left(2\pi\frac{t}{T}\right)\right].
\end{split}
\end{equation}
%----
We then rotate $(\xi,\zeta)$ about the $z$-axis by an angle {$\lambda-\mathrm{arctan}\left({\xi(l,t)}/{\zeta(l,t)}\right)$} to produce the cilium bending motion in the $(x,y)$ plane, see Fig.~\ref{fig:model}B. 
To make the cilium swing out of the vertical plane, we perform a second rotation about the $x$-axis by the angle $\beta$, see Fig.~\ref{fig:model}C. 
At the end of  these two rotations, one obtains cilia-like kinematics $(x_c(s,t), y_c(s,t), z_c(s,t))$ that satisfy the desired design parameters $(\lambda_o, \alpha_o, \beta_o)$. This family of beating kinematics qualitatively resembles the beating patterns of biological cilia in that the effective stroke is straight and perpendicular to the base surface while the recovery stroke is curly and close to the base surface. The beating kinematics, by construction, spend equal amount of time in effective and recovery strokes. The hydrodynamic performance, however, is independent of time because the flow in Stokes regime is essentially a geometric problem. 
%\ek{Hanliang: this paragraph is very ambiguous. There is no way anyone will understand what you doing here if we don't do a better job at explaining it. Do you start by choosing $\lambda_o$, $\beta_o$ and $\alpha_o$, then use (2) to reconstruct $(x,y,z)$? if so, how exactly?}

%In general, the recovery stroke ($\frac{1}{2}\le \phi 1$) of the generic beating pattern has a larger curvature than the effective stroke ($\0\le \phi \frac{1}{2}$) and lies closer to the base wall.

We use $l$ and $T$ to scale length and time, respectively.  All variables are thereafter non-dimensional.
The fluid motion is governed by the non-dimensional Stokes equation and incompressibility condition at zero Reynolds number,
%----
\begin{equation}	\label{stokesequation}
-\nabla p + \mu\nabla^2\mathbf{u} = \mathbf{0},\qquad \nabla\cdot\mathbf{u} = 0,
\end{equation}
%----
where $p$ is the pressure field, $\mu$ is the dimensionless viscosity, and $\mathbf{u}$ is the fluid velocity field. The boundary condition at the cilium and base surface is  a no-slip condition
%------
\begin{equation}	\label{bc}
\begin{split}
\mathbf{u}|_{\mathrm{boundary}}&= \left\{ \begin{array}{l l}
	\mathbf{u}_c \qquad \text{at the cilium} \\
	\mathbf{0} \ \qquad\text{at the base surface}
	\end{array} \right. , % \qquad \qquad \mathbf{u}|_{\mathrm{\infty}} = 0.
\end{split}
\end{equation}
%-------
where $\mathbf{u}_c$ is the prescribed velocity of the cilium, based  on  the  kinematics  reconstructed either from experimental images or the mathematical formulation.

We solve (\ref{stokesequation}) and (\ref{bc}) numerically using the regularized Stokeslet method \cite{cortez2005}. The cilium is approximated by a distribution of regularized Stokeslets along its centerline, together with an ``image'' distribution to impose the no-slip boundary conditions at the base surface~\cite{ainley2008}.
%The velocity at an arbitrary point x in the fluid domain is approximated by the sum over all Stokeslets (i = 1, . . . , $n$, where $n$ is the total number of Stokeslets)
%where $\mathbf{G}_s(\mathbf{x}-\mathbf{x}_i)\cdot\mathbf{f}_i$ is the velocity at x induced by a regularized Stokeslet of strength $\mathbf{f}_i$ located at $\mathbf{x}_i$.
The velocity at any point $\mathbf{x}$ induced by a regularized Stokeslet of strength $\mathbf{f}_i$ located at $\mathbf{x}_i$ and its images can be written as $\mathbf{G}_s(\mathbf{x}-\mathbf{x}_i)\cdot\mathbf{f}_i$, where $\mathbf{G}_s$ is the regularized Green's tensor-valued function given by Ainley \textit{et al.}~\cite{ainley2008}. The total velocity generated by all regularized Stokeslets is then
%----------
\begin{equation}	\label{regstoke}
\mathbf{u}(\mathbf{x}) = \sum_{i=1}^{n}\mathbf{G}_s(\mathbf{x}-\mathbf{x}_i)\cdot\mathbf{f}_i,
\end{equation}
%----------
where $n$ is the total number of regularized Stokeslets. The expression in~\eqref{regstoke} is substituted into~\eqref{bc} to obtain a system of equations that can be used to compute  the strengths $\mathbf{f}_i$ of the regularized Stokeslets.
% can be solved from equation \eqref{regstoke} together with the prescribed velocities and no-slip boundary conditions at the cilium. 
Once  $\mathbf{f}_i$ are known, the flow field can be reconstructed everywhere.

Let $Q_x$ and $Q_z$ denote the flow transported by the cilium in the $x$- and $z$-directions, respectively. 
%By virtue of incompressibility, the flow transported by the cilium in the $x$-direction is equal to the flow across any half plane given by   $x=$constant. Similarly, the flow transported in the $z$-direction is equal to the flow across any half plane given by  $z=$ constant.  
%The flow rates $Q_x$ and $Q_z$  in the $x$- and $z$-directions, respectively,  are given by
One has (see~\cite{liron1978fluid,smith2008,ding2014} for more details) 
%----
\begin{equation}
Q_x = \frac{1}{\mu\pi}\int_0^l(\mathbf{f}_i\cdot\mathbf{e}_x)yds, \qquad Q_z = \frac{1}{\mu\pi}\int_0^l(\mathbf{f}_i\cdot\mathbf{e}_z)yds
\end{equation}
%---
%\eknote{add expression for $Q_z$ in equation (7)}
 where $\mathbf{e}_x$ and $\mathbf{e}_z$ are the unit vectors  in $x$- and $z$-directions. {These expressions are based on the fact that the net flux generated by a Stokeslet over an infinite plane is directly proportional to the height of the Stokeslet above the plane~\cite{liron1978fluid}. Now, consider the time average of the flow rates per cycle:} $\langle Q_x\rangle = \frac{1}{T}\int_0^TQ_xdt$ and $\langle Q_z\rangle = \frac{1}{T}\int_0^TQ_zdt$. The total average flow rate is $\langle Q\rangle = \sqrt{\langle Q_x\rangle^2 + \langle Q_z\rangle^2}$.

To compute the internal power spent by the cilium during the beating cycle, we  consider each cilium to be an inextensible elastic filament and adopt the Kirchhoff equations of motion for a rod~\cite{eloy2012, guo2014} 
%---
\begin{equation}	\label{eq:rod}
\frac{\partial \mathbf{N}}{\partial s}-\mathbf{f} = 0, \qquad \frac{\partial \mathbf{M}}{\partial s}+\mathbf{t}\times\mathbf{N}+\mathbf{q} = 0.
\end{equation}
%---
Here %$s$ is the arc-length measured from the root of the cilium, 
$\mathbf{N}(s,t)$ and $\mathbf{M}(s,t)$ are the internal tension and elastic moment respectively, $\mathbf{f}(s,t)$ is the force exerted by the cilium on the surrounding fluid, $\mathbf{t} = \partial \mathbf{x}_c/\partial s$ is the unit tangent to the cilium, and $\mathbf{q}(s,t)$ is the internally generated moment per unit length. The internal bending moment $\mathbf{q}$ resembles the moments generated by the cilium internal motors.

The force distribution $\mathbf{f}$ is computed by dividing the local Stokeslet strength $\mathbf{f}_i$ by the distance between neighboring Stokeslets along the cilium. We assume  a linear constitutive relation between the elastic moment $\mathbf{M}$ along the cilium and the deformation~\cite{eloy2012,guo2014}. Namely, we let $\mathbf{M} = B\mathbf{D}$, where $B$ is the bending rigidity and $\mathbf{D}=\mathbf{t}\times\left({\partial\mathbf{t}}/{\partial s}\right)$ is the Darboux vector. Substituting $\mathbf{M} = B\mathbf{D}$  into~\eqref{eq:rod}, one gets the expression of the internal moments generated along the cilium 
%-----
\begin{equation}
\mathbf{q} = B\frac{\partial^2 \mathbf{t}}{\partial s^2}\times\mathbf{t}\ +\ \mathbf{t}\times\!\!\int_s^l\mathbf{f}(\tilde{s},t)\text{d}\tilde{s}.
\end{equation}
%-----
The average power $\langle P \rangle$ expended internally by the cilium to transport is equal to the power consumed by the internal moments $\mathbf{q}$, 
%----
\begin{equation}
\langle P \rangle = \left\langle \int_0^l \max(0,\mathbf{q}\cdot\boldsymbol{\Omega})\text{d}s\right\rangle,
\end{equation}
%---
where $\boldsymbol\Omega = ||\mathbf{\dot{t}}(s)||\frac{\mathbf{t}\times\mathbf{\dot{t}}}{||\mathbf{t}\times\mathbf{\dot{t}}||}$ is the angular velocity vector, {with $\dot{( )}$ denoting the time derivative}. Here,  only the positive works are accounted for, i.e., the cilium does not harvest energy from the environment~\cite{machin1958wave, eloy2012}. This implies that the mean power spent by the cilium's internal moments is larger than the power given to the fluid.
%Note that accounting for only positive work means that the dynein arms in the cilium cannot harvest energy from the fluid environment and implies that the mean power spent by the internal moments is larger than the power given to the fluid8.
%The negative works are not taken into account as the cilium cannot harvest energy from the ambient.
Finally, we define a dimensionless transport efficiency,
\begin{equation}
\eta = \mu l^{-3}\frac{\langle Q\rangle^2}{\langle P \rangle},
\end{equation}
which is consistent with that employed by Osterman \& Vilfan~\cite{osterman2011} and Eloy \& Lauga~\cite{eloy2012}.

%--------------------
\begin{figure*}[t]
\begin{center}
{\includegraphics{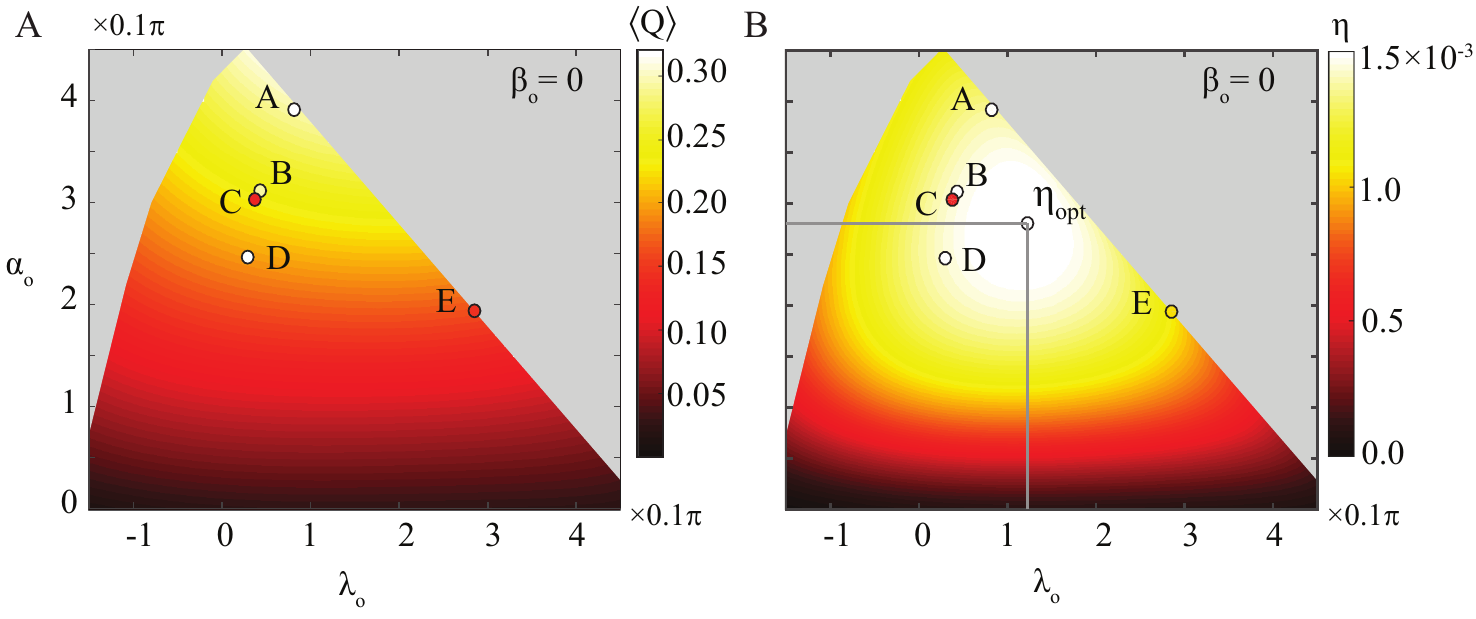}}
\end{center}
  	\caption[]{Flow rate $\langle Q \rangle$ and efficiency $\eta$ generated by the generic cilia model with $\beta_o = 0$. Flow rates and the efficiencies of the  cilia beating patterns reconstructed in Fig.~\ref{fig:strokes} are plotted in circles at corresponding values of $\lambda_o$ and $\alpha_o$. The color of the circle indicates the performance of these cilia using the colormap of each subfigure. Deviations between the background color and the circle color indicate deviations in the performance of the generic model and  actual cilia.}
	\label{fig:comparison}
\end{figure*}
%------------------------

\section{Results}
\label{sec:results}

Our goal is to evaluate the hydrodynamic performance of the beating kinematics of individual cilia as a function of the three design parameters: the leaning angle $\lambda_o$, the beating amplitude $\alpha_o$ and the out-of-plane swinging angle $\beta_o$. We use two evaluation metrics: the average flow rate $\langle Q\rangle$ and the efficiency $\eta$. 

 We discretize the cilium uniformly using $20$ regularized Stokeslets. The regularization parameter is chosen to be $0.05$, which yields a cilium length-to-radius ratio about $20$. Each beating cycle is discretized into $100$ time steps. By way of validation, we are able to reproduce the flow rates generated by embryonic primary cilia given by Smith \textit{et al.}~\cite{smith2008} with these discretization parameters. 
 %\ek{this is a very confusing remark. Why should the reader care that we are able to reproduce these flow rates? is it by way of validation? you should state it explicitly then!}

We begin by considering planar cilia beat kinematics, for which $\beta_o=0$. We examine the performance of the cilium as a function of $\lambda_o$ and $\alpha_o$. Namely, we vary $\lambda_o$  in the range from $-0.15\pi$ to $0.45\pi$ and we vary $\alpha_o$ from $0$ to $0.45\pi$ using a step size of $0.01\pi$. Combinations of parameters for which $\lambda_o + \alpha_o>0.5\pi$ lead to prescribed cilium motion that penetrate the base surtface. These  combinations are considered unrealistic and not considered. 
Note that accurate information regarding the angular frequency and bending rigidity of cilia is sparse. Here, as well as in the rest of this section, 
%Given the sparse information about the angular frequency and bending rigidity of cilia, 
we use the information available for the {\em Paramecium} as a proxy
to obtain the right order of magnitude for the cilia angular frequency and bending rigidity.  The typical length of the {\em Paramecium} cilia is about $10\mu \text{m}$, typical angular frequency  is about $200 \text{rad}\cdot\text{s}^{-1}$, and the bending rigidity is estimated to be $B=25\text{pN}\cdot\mu\text{m}^2$ (see~\cite{brennen1977,hines1983,eloy2012}). We thus use the characteristic length $l_c = 10\mu \text{m}$, time $T_c = 2\pi/200 = 0.0314\text{s}$, viscosity $\mu_c = 10^{-3}\text{Pa}\cdot\text{s}$. The non-dimensional bending rigidity is  then given by $B/(\mu_c l_c^4T_c^{-1}) = 0.0785$. \textcolor{black}{Note that more recent estimates of the bending rigidity predict higher values of $B$,~e.g.,~\cite{gittes1993flexural}. However, the exact $B$ value does not affect the main findings of this work, which focuses mainly on a comparative analysis of cilia performance under various design parameters.}

%because we are comparing beating patterns whose shape is not sensitive to $B$. rather than.  between generic beating patterns with different design parameters are not sensitive to the estimated bending rigidity because the differences across these patterns are cilium orientations only, not shapes.}

%--------------------
\begin{figure*}[t]
\begin{center}
%{\includegraphics[width=\linewidth]{figs/Q_Eta_layers_panelV4.eps}}
{\includegraphics[width=\linewidth]{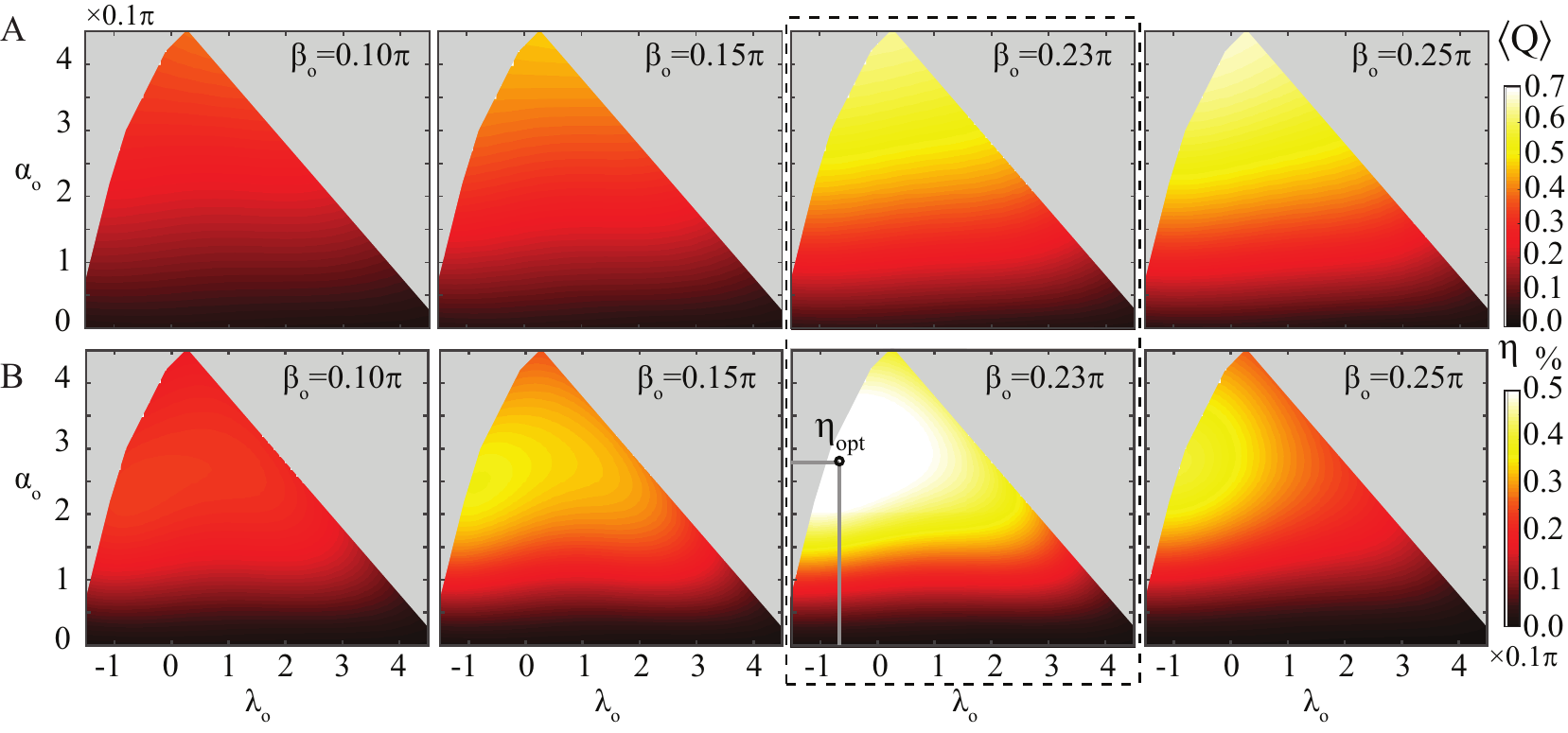}}
\end{center}
  	\caption[]{A: average flow rates $\langle Q \rangle$ of generic cilia kinematica with different beating amplitude $\alpha_o$, leaning angle $\lambda_o$, and swinging angle $\beta_o$. B: Transport efficiencies.}
	\label{fig:QPanel}
\end{figure*}
%------------------------

The average flow rate $\langle Q\rangle$ and transport efficiency $\eta$ for the admissible parameter values $(\lambda_o,\alpha_o)$  are shown in Fig.~\ref{fig:comparison}.
It is clear from Fig.~\ref{fig:comparison}A that larger beating amplitude $\alpha_o$ lead to larger average flow rate. Meanwhile the leaning angle $\lambda_o$ has a small effect on the average flow rate. The beating patterns with {small} beating amplitude {($\alpha_o \approx 0$)} correspond to traveling waves {somewhat similar} to those observed in the flagella of sperm cells~\cite{brennen1977,lauga2009}. {However, unlike the traveling waves in natural sperm cells, these waves are constrained such that, the head and tip of the wave always lie at a constant direction from the substrate and the curvature of the wave has the same sign. If one were to use a flagellated sperm cell and adhere its head to a substrate such that its flagellum beats {in such constraint traveling wave fashion}, it would produce little flow.}

The effect of $\alpha_o$ and $\lambda_o$ on the transport efficiency $\eta$ is depicted in Fig.~\ref{fig:comparison}B.  The beating pattern with $\lambda_o = 0.12\pi$ and $\alpha_o = 0.28\pi$ yields the highest transport efficiency $0.15\%$. Unlike its effect on $\langle Q \rangle$, a larger beating amplitude does not yield higher efficiency. This is 
because at higher $\alpha_o$  the cilium will need to spend more power in its motion close to the base surface to overcome the zero velocity at that surface. The extra power needed to complete such beating cycles lowers the transport efficiency.

By way of validation of these generic cilia-like kinematics, we compute the performance of the five beating patterns  reconstructed from experimental data shown in Fig.~\ref{fig:strokes}. The leaning angles and beating amplitudes of these cilia are calculated according to~\eqref{eq:parameters}. The flow rates and the transport efficiencies are then computed  and the results are superimposed onto Fig.~\ref{fig:comparison} as colored circles in accordance with their leaning angles and beating amplitudes. Clearly, the experimentally-derived beating patterns are scattered in the high efficiency zone predicted by the generic model. Furthermore, the flow rates and the transport efficiencies predicted by the generic kinematics are close to those obtained from experimental data, except for one data point which differs from the generic model by a factor of about 2. These findings justify our choice of both the reduced design parameters and the generic model. 

\begin{figure*}[t]
\begin{center}
{\includegraphics[width=\textwidth]{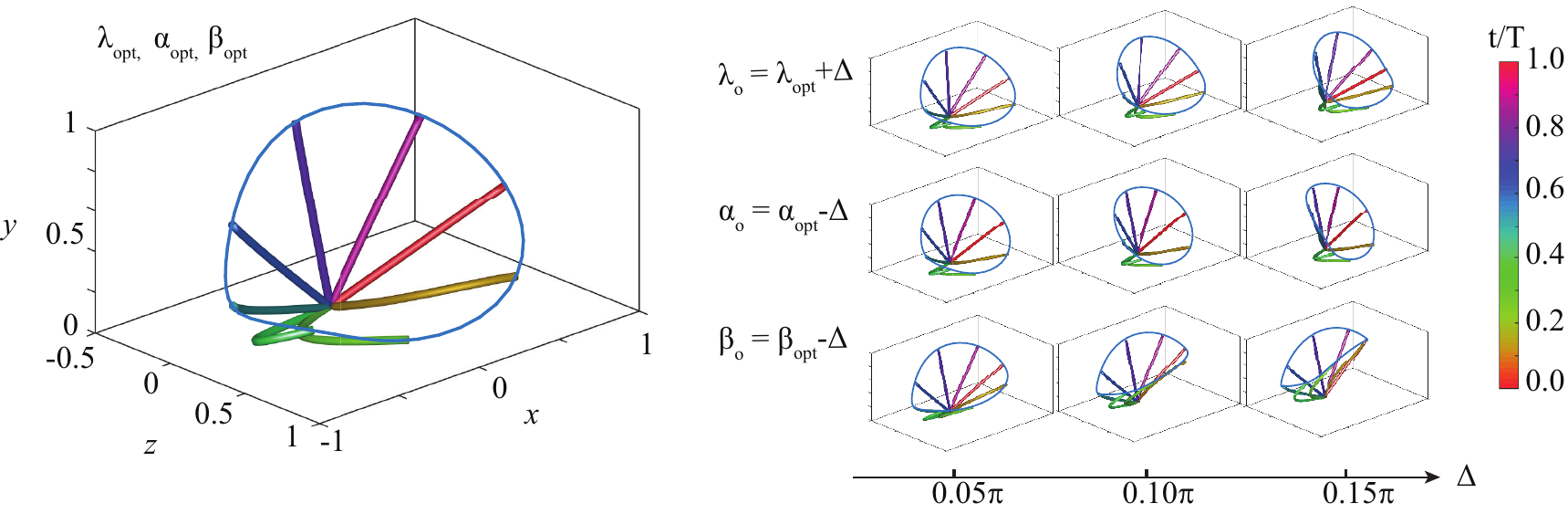}}
\end{center}
  	\caption[]{Optimal (left) and suboptimal (right) beating patterns. The suboptimal beating patterns are constructed by perturbing one of the three design parameters while keeping the other two the same as in the optimal beating pattern.
}
	\label{fig:sensepanel}
\end{figure*}
%------------------------

%--------------------
\begin{figure*}[t]
\begin{center}
{\includegraphics{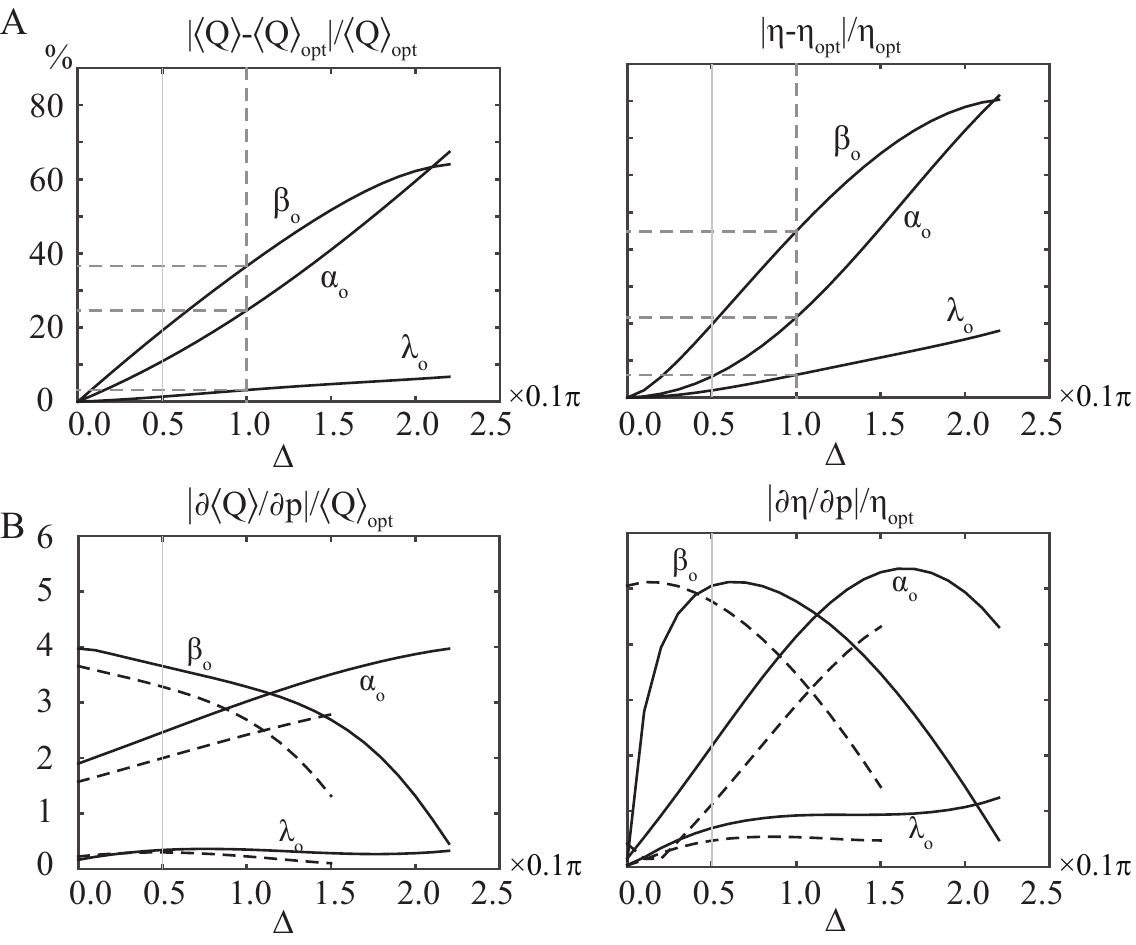}}
\end{center}
  	\caption[]{Sensitivity of the net flow rate $\langle Q \rangle$ and the transport efficiency $\eta$ with respect to perturbations $\Delta$ away from the optimal design parameters $\alpha_{\rm opt}$, $\lambda_{\rm opt}$ and $\beta_{\rm opt}$. A:  ``global'' sensitivities or net change in $\langle Q \rangle$ and $\eta$ as a function of $\Delta$. B: ``local''  sensitivities or rate of change in $\langle Q \rangle$ and $\eta$ as a function of perturbations $\Delta$ away from optimal parameter values (solid lines) and away from suboptimal parameter values (dashed lines) .
}
	\label{fig:sensitivity}
\end{figure*}
%------------------------

%--------------------
\begin{figure*}[t]
\begin{center}
{\includegraphics{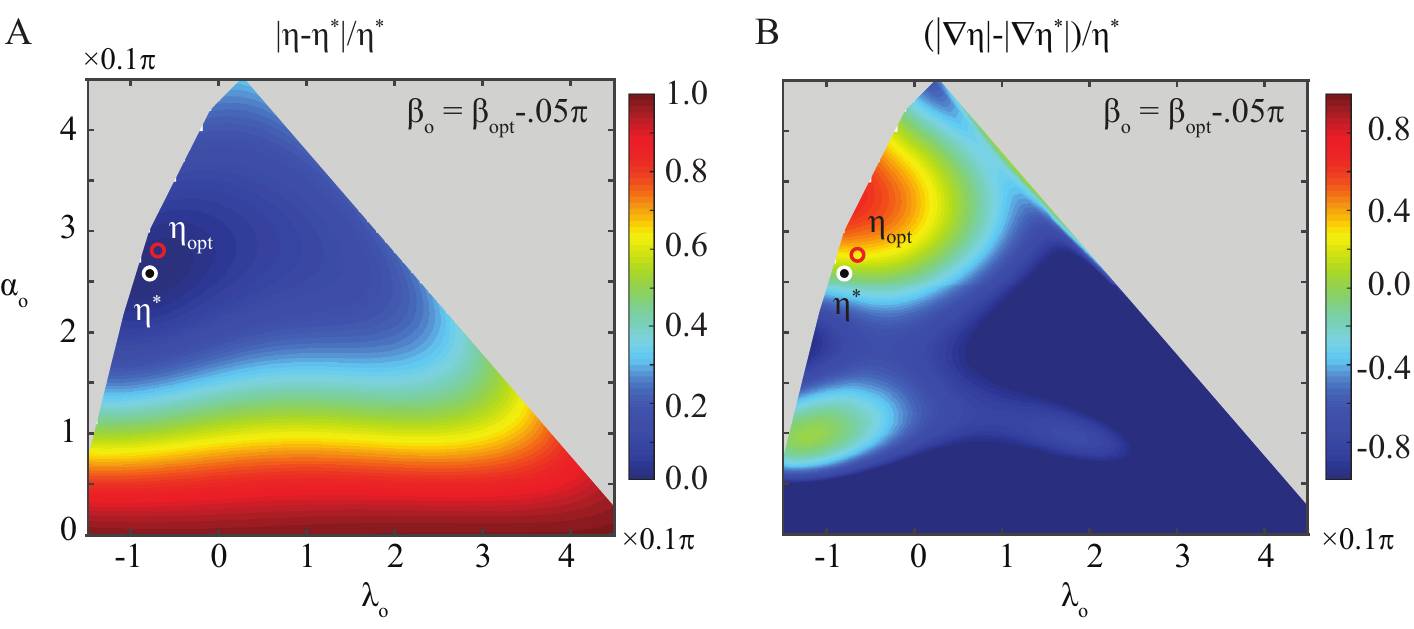}}
%{\includegraphics{Fig8V2.eps}}
\end{center}
  	\caption[]{``global'' (A) and ``local'' (B) sensitivities of the transport efficiency measured from a suboptimal choice of $\beta^\ast$ as  functions of $\lambda_o$ and $\alpha_o$ for $\Delta = 0.05\pi$.
}
	\label{fig:sens_contour}
\end{figure*}
%------------------------

We now consider three-dimensional cilia kinematics and assess the effects of the out-of-plane swinging angle $\beta_o$ on $\langle Q\rangle$ and $\eta$. Namely, we vary $\beta_o$ from zero (results in Fig.~\ref{fig:comparison}) to $0.25\pi$ using increments of $0.01\pi$. In Figs.~\ref{fig:QPanel}, the values of $\langle Q\rangle$ and $\eta$ are depicted for select values of $\beta_o$. 
Fig.~\ref{fig:QPanel}A shows that a larger  swinging angle $\beta_o$ leads to a larger average flow rate. This is similar to the effect of the beating amplitude $\alpha_o$ on the flow rate:  the largest values of $\alpha_o$ and $\beta_o$ ($\alpha_o = 0.45\pi$, $\beta_o = 0.25\pi$) yields the largest average flow rate. 
%The flow rates are less sensitive to leaning angles for a given beating amplitude and swinging angle, compare to the other two parameters.  
%This is understandable as the flow dragged back by the recovery stroke is limited by the presence of the wall when $\beta_o$ is close to $0.25\pi$, in which case the recovery stroke is moving close to the wall.
%When $\beta_o$ is close to $0.25\pi$, the recovery stroke would be moving close to the base wall by construction. Hence a beating pattern with large $\beta_o$ understandably generates large flow rate over one cycle, as the flow dragged back during the recovery stroke is limited by the presence of the wall. 
The efficiency, on the other hand, is not a monotonically increasing function of $\beta_o$. Fig.~\ref{fig:QPanel}B shows that the efficiencies for $\beta_o = 0.23\pi$ are higher than those for $\beta_o = 0.25\pi$ at the same values of $\lambda_o$ and $\alpha_o$. In fact, we chose to depict the panels for $\beta_o = 0.23$ because it is the out-of-plane angle that leads to maximum transport efficiency for $\lambda_o = \lambda_{\rm opt} = -0.07\pi$, $\alpha_o= \alpha_{\rm opt} = 0.28\pi$. %and $\beta_o = \beta_{\rm opt} = 0.23\pi$ 
The maximum efficiency value is $0.56\%$. 
%As we discussed earlier for large $\alpha_o$, moving close to the base wall would require significantly higher power, hence limiting the transport efficiencies in such scenarios. 

%%--------------------
%\begin{figure*}[t]
%\begin{center}
%%{\includegraphics[width=\linewidth]{figs/Eta_layers_panelV5.eps}}
%\end{center}
%  	\caption[]{Transport efficiencies of generic beating patterns with different beating amplitude $\alpha_o$, leaning angle $\lambda_o$, and swinging angle $\beta_o$.}
%	\label{fig:EffPanel}
%\end{figure*}
%%------------------------

To study the sensitivity of the cilia performance with respect to the design parameters $(\lambda_o,\alpha_o,\beta_o)$, we perturb each one of these parameters away from the optimal or most efficient combination while keeping the other two parameters the same. In particular, we let  $\lambda_o = \lambda_{\rm opt}+\Delta$, $\alpha_o = \alpha_{\rm opt}-\Delta$, and $\beta_o = \beta_{\rm opt}-\Delta$, where $(\lambda_{\rm opt},\alpha_{\rm opt},\beta_{\rm opt})$ is the optimal parameter combination that leads to the most efficient cilium. {The most efficient beating pattern and the beating patterns perturbed from the most efficient one are shown in Fig.~\ref{fig:sensepanel}.} %so that we can explore a wide range of $\Delta$.

We introduce two measures of sensitivity: (i) a ``global'' sensitivity which evaluates the net change in flow transport
{$\left|\langle Q \rangle - \langle Q_{\rm opt} \rangle \right|/\langle Q_{\rm opt} \rangle$}, and (ii) a ``local'' sensitivity which evaluates the rate of change in flow transport normalized by $\langle Q_{\rm opt} \rangle$. Here, by rates of change in flow transport we mean  $\left|\partial \langle Q \rangle /\partial \lambda_o\right|$, $\left|\partial \langle Q \rangle /\partial \alpha_o\right|$, and $\left|\partial \langle Q \rangle /\partial \beta_o\right|$. Similarly, the {net change in} efficiency {$\left|\eta-\eta_{\mathrm{opt}}\right|/\eta_{\mathrm{opt}}$} and rates of change of efficiency  $\left|\partial \eta  /\partial \lambda_o\right|$, $\left|\partial \eta  /\partial \alpha_o\right|$,  and $\left|\partial \eta  /\partial \beta_o\right|$ are, respectively, ``global'' and ``local'' measures of the sensitivity in cilia efficiency as the beating kinematics deviate from that of the most efficient cilium. A lower sensitivity implies more robustness to perturbations imposed on cilia design parameters, and vice versa.

In Fig.~\ref{fig:sensitivity}A is a depiction of the global sensitivities of both the  transport flow rate and efficiency as a function of the deviation $\Delta$ from the optimal cilium kinematics. 
Both the flow rate and efficiency are more sensitive to the beating amplitude $\alpha_o$ and out-of-plane swinging amplitude $\beta_o$ than to the leaning angle $\lambda_o$. For example, a perturbation $\Delta = 0.1\pi$ {(dashed grey lines in Fig.~\ref{fig:sensitivity}A)} in the swinging angle $\beta_o$ will reduce the flow rate by over $36\%$, as compared to a reduction of $25\%$ and $3\%$ when varying the beating amplitude $\alpha_o$ and leaning angle $\lambda_o$ by the same amount, respectively. A similar trend is observed for the sensitivity in the transport efficiency $\eta$. The local sensitivity measures, shown in Fig.~\ref{fig:sensitivity}B confirm these findings but indicate that for larger deviations from the optimal kinematics, the  performance of the cilium becomes more sensitive to variations in beating amplitude $\alpha_o$. These results indicate that to maintain lower sensitivity at large perturbations away from the optimal beating kinematics, a better strategy is to allow deviations/variations in $\beta_o$  while restricting the variations in $\alpha_o$. %\ek{why is that important?} 
In other words, ciliary defects that induce large variations in $\alpha_o$ lead to  ineffective flow transport.

Fig.~\ref{fig:sensitivity}B also shows the sensitivities  associated with a suboptimal choice of cilia design parameters (dashed lines). Namely, instead of perturbing the parameters from $\left( \lambda_{\mathrm{opt}}, \alpha_{\mathrm{opt}}, \beta_{\mathrm{opt}}\right)$, we consider the suboptimal parameters $\left( \lambda_{\mathrm{opt}}, \alpha_{\mathrm{opt}}, \beta^*\right)$, where $\beta^* = \beta_{\mathrm{opt}}-0.05\pi$ and impose perturbations $\Delta$ away from this suboptimal choice.   Our goal is to compare the 
sensitivity of optimal and suboptimal design: the design with lower sensitivity is more robust to perturbations. It is evident from
Fig.~\ref{fig:sensitivity}B that the ``local'' sensitivities of the flow rate and efficiency associated with the suboptimal design  are lower than those associated from the optimal design for almost all $\Delta$, indicating that a suboptimal design may lead to more robust performance when the design parameters are perturbed. Note that the only sensitivity  component that is increased with this suboptimal choice of parameters is the sensitivity of transport efficiency with respect to the out-of-plane angle $\beta_o$ when the perturbation $\Delta$ is very small $\Delta < 0.04\pi$.

Finally, we set $\Delta = 0.05\pi$ (highlighted in light grey in Fig.~\ref{fig:sensitivity}) and we calculate the global and local sensitivities in transport efficiency $\eta$ for all admissible values of $\lambda_o$ and $\alpha_o$ for $\beta_o = \beta^\ast = \beta_{\rm opt} - \Delta$. 
In particular, we let $(\lambda^*,\alpha^*)$ denote the parameters that lead the highest transport efficiency $\eta^*$ for $\beta_o = \beta^*$, and we measure the global and local sensitivities relative to $\eta^\ast$.
%. is depicted in the solid black circle, while $(\lambda_{\mathrm{opt}},\alpha_{\mathrm{opt}})$ is depicted in the red circle. 
The results are depicted in Fig.~\ref{fig:sens_contour}A and B, respectively. In the latter, the  local sensitivities with respect to the design parameters $\alpha_o$, $\lambda_o$ and $\beta_o$ are consolidated  into one scalar function {$(\left| \nabla\eta\right| - \left| \nabla\eta^*\right|)/\eta^*$}, where $\nabla\eta$ is defined as $\nabla \eta = \sqrt{(\partial\eta/\partial\lambda_o)^2+(\partial\eta/\partial\alpha_o)^2+(\partial\eta/\partial\beta_o)^2}$. Such depiction serves as a tool to identify the design parameters $\alpha_o,\lambda_o$ that satisfy desired limits on sensitivity given a perturbation size, in this case $\Delta = 0.05\pi$. For example, to ensure minimum sensitivities and maximum robustness for this perturbation size, one would choose values of $\alpha_o$ and $\lambda_o$ that lie in the intersection of the regions \textcolor{black}{that correspond to minimal sensitivities} in Fig.~\ref{fig:sens_contour}A and B.

{A few comments on the interplay between efficiency and robustness are in order. At  optimal efficiency, one has $\nabla\eta =0$ by definition, which implies infinite robustness. However,  cilia are often subject to perturbations in the fluid environment as well as in the ciliary apparatus. But any perturbation away from the optimal beating pattern causes  the cilium to loose efficiency.   It is thus reasonable to assume that cilia often operate at suboptimal efficiency. The question then is, starting from  suboptimal efficiency, how is cilia transport affected by further perturbations to the cilia beating kinematics. Interestingly, we found that starting from suboptimal efficiency, cilia transport may be more robust to perturbations than starting from optimal efficiency, hence, the most efficient cilium is not the most robust.}

\section{Discussion}

We parameterized cilia beating kinematics using three reduced design parameters:  the leaning angle $\lambda_o$ of the cilium in the direction of the effective stroke, the cilium beating amplitude $\alpha_o$, and its out-of-plane swinging angle $\beta_o$. We presented a straightforward  method for extracting these three parameters  from experimental data. We also introduced a mathematical approach based on a sinusoidal family of time-dependent traveling waves that generate  ``generic'' cilia-like kinematics with prescribed values of $\lambda_o, \alpha_o$ and $\beta_o$. To examine the performance of the various cilia kinematics, those obtained from experimental image sequences and those generated mathematically, we  used two performance metrics: the net flow transported  by the cilium and its efficiency. 

We compared the performance of the ``generic" cilia-like kinematics (Fig.~\ref{fig:model}) with those reconstructed from experiments (Fig.~\ref{fig:strokes}). We restricted this comparison to the case of planar beating cilia due to the lack of experimental data that accurately resolve the out-of-plane cilia beating kinematics. In the planar case, one has only two design parameters: the leaning angle $\lambda_o$ and beating amplitude $\alpha_o$.  The flow rates and the transport efficiencies predicted by the generic kinematics are of the same order of magnitude as  those obtained from experimental data (Fig.~\ref{fig:comparison}). Further, the design parameters extracted from the experimental data are all located in the region of the parameter space where the transport efficiencies are highest.
%Although the generic beating pattern could not perfectly predict the actual flow rates and the transport efficiencies of the real beating patterns, we found the design parameters of the real beating patterns located in the parameter domain where the transport efficiencies were high according to our results. 
Together, these observations provide strong evidence that both the reduced design parameters and the generic cilia-like kinematics proposed in this work capture the salient features of cilia beating patterns.

%The goal of this work is to provide a quantitative basis for comparing the performance of cilia from different cell type, as well as  cilia from the same cell type under different operating conditions such as in healthy and diseased states. To this end, we parameterized the cilia beating kinematics using three reduced design parameters:  the leaning angle $\lambda_o$ of the cilium in the direction of the effective stroke, the cilium beating amplitude $\alpha_o$, and the out-of-plane swinging angle $\beta_o$. We presented a straightforward  method for extracting these three parameters  from experimental data. We also introduced a  ``generic'' method for generating cilia-like kinematics with prescribed values of $\lambda_o, \alpha_o$ and $\beta_o$ based on a sinusoidal family of time-dependent traveling waves. To examine the performance of the individual cilia, those obtained from experimental image sequences and those generated mathematically, we  used two performance metrics: the net flow transported  by the cilium and its efficiency. 

We then presented a systematic study of the two performance metrics as a function of $\lambda_o$, $\alpha_o$, $\beta_o$ in the context of the generic cilia model. Our results suggest that the flow rates are positively affected by the beating amplitude $\alpha_o$ and the out-of-plane swinging angle $\beta_o$ in the sense that  larger values of $\alpha_o$ and/or $\beta_o$ generate  larger flow rates (Fig.~\ref{fig:QPanel}). Indeed, both parameters amplify the asymmetry between the effective and recovery stroke, thus enhancing flow transport.
Meanwhile, the transport efficiency  is maximum for an optimal set of parameter values. Values $\alpha_o$ and $\beta_o$ larger than these optimal values result in the cilium moving closer to the no-slip base surface. Since larger internal power  is needed to move the cilium  closer to the no-slip surface, such values lead to lower transport efficiency. 
%Specifically, the design parameters of the most efficient beating patterns were $\lambda_o = -0.07\pi$, $\alpha_o = 0.28\pi$, and $\beta_o = 0.23\pi$ for the generic beating pattern, yielded a transport efficiency of $0.56\%$.

Finally, we examined the sensitivity of the flow rate and efficiency with respect to the design parameters. We investigated by how much the flow rate and efficiency would change as the design parameters deviate from the most efficient parameters. We called this a ``global'' sensitivity. We also examined a ``local'' sensitivity to parameter changes by computing how `fast' the transport flow rate and efficiency would change when perturbed away from the most efficient cilium.  Our results show that perturbations in the out-of-plane angle $\beta_o$  or beating amplitude $\alpha_o$ induce large sensitivities, that is, a deterioration in ciliary performance, whereas
%when  compared to a similar perturbation in the other two parameters if the perturbation size is larger than $0.2\pi$. Otherwise, a perturbation in the beating amplitude $\alpha_o$ has a more significant effect. %as the change rate of the transport performances with respect to the beating amplitude surpassed the change rate with respect to the swinging angle when $\Delta>0.11\pi$. 
a change of the leaning angle $\lambda_o$ has no notable effect on the cilia performance (Fig.~\ref{fig:sensitivity}). Table~\ref{tab:perturb} summarizes  and compares the perturbation size needed in each of the design parameters to induce a 10\% drop in the ciliary performance. 
%Our results showed that the transport performances were most sensitive to the swinging angle when the perturbation is relatively small ($\Delta<0.11\pi$); the beating amplitude would become the most sensitive parameter when the perturbation is larger. The change of the leaning angle, on the other hand, did not affect the transport performance significantly.

%-------------------
 \begin{table}[!h]	
  \caption{ \label{tab:perturb} Perturbations of the design parameters $(\lambda_o,\alpha_o, \beta_o)$ which lead to a  $10\%$ drop in the net flow rates and transport efficiencies relative to the optimal values.}
 %\begin{adjustwidth}{-2in}{-2in}
 \centering
 {\footnotesize
 \begin{tabular}{c |c| c c c|}
 \cline{2-5}
  &   & $\lambda_o$ & $\alpha_o$  & $\beta_o$  \\
 \cline{2-5}
& $\langle Q\rangle$ & $-$ & $0.05\pi$ &$0.03\pi$ \\
& $\eta$ & $0.12\pi$ &$0.06\pi$ &$0.03\pi$ \\
 \cline{2-5}
 \end{tabular} }
 %\end{adjustwidth}
 %\vspace{-1.25\baselineskip}
 \end{table}
 %-------------------

Most importantly, our sensitivity analysis shows that the most efficient cilium is not most robust to perturbations in the cilia kinematics. Indeed, we presented a counterexample showing suboptimal parameters that are more robust (less sensitive) to perturbations (Fig.~\ref{fig:sens_contour}). Our results have two major implications. First, they confirm that designing for the most efficient cilium does not automatically impose any guarantees on robustness. It is our view that from the standpoint of evolutionary biology, robustness of design to natural or acquired variations is as important as efficiency itself. Therefore, when using genetic computational algorithms to compute ciliary design~\cite{eloy2012,osterman2011},  one has to explicitly account for robustness. Second, our methods and results provide a quantitative framework for comparing the performance of cilia from different cell type, as well as cilia from the same cell type under different operating conditions such as in healthy and diseased states.  To this end, one can begin to investigate quantitatively how perturbations and/or disruptions of the ciliary apparatus, whether due to a genetic disorder
or infective and acquired causes,  affect the flow transport. Low and inefficient flow rates of ciliated surfaces in mammalian organisms are directly linked to infection and disease, such as in cystic fibrosis and asthma. A quantitative map from cilia parameters to flow rates and vice versa would therefore provide an important tool for assessing cilia performance in health and disease. 
%Finally, we conjecture that robustness to environmental and inherent variations may play a role in explaining the diverse cilia kinematics reported experimentally. 

\begin{acknowledgments}
Computation for the work described in this paper was supported by the University of Southern California's Center for High-Performance Computing (http://hpcc.usc.edu).
\end{acknowledgments}

\bibliography{references}

\begin{thebibliography}{36}%
\makeatletter
\providecommand \@ifxundefined [1]{%
 \@ifx{#1\undefined}
}%
\providecommand \@ifnum [1]{%
 \ifnum #1\expandafter \@firstoftwo
 \else \expandafter \@secondoftwo
 \fi
}%
\providecommand \@ifx [1]{%
 \ifx #1\expandafter \@firstoftwo
 \else \expandafter \@secondoftwo
 \fi
}%
\providecommand \natexlab [1]{#1}%
\providecommand \enquote  [1]{``#1''}%
\providecommand \bibnamefont  [1]{#1}%
\providecommand \bibfnamefont [1]{#1}%
\providecommand \citenamefont [1]{#1}%
\providecommand \href@noop [0]{\@secondoftwo}%
\providecommand \href [0]{\begingroup \@sanitize@url \@href}%
\providecommand \@href[1]{\@@startlink{#1}\@@href}%
\providecommand \@@href[1]{\endgroup#1\@@endlink}%
\providecommand \@sanitize@url [0]{\catcode `\\12\catcode `\$12\catcode
  `\&12\catcode `\#12\catcode `\^12\catcode `\_12\catcode `\%12\relax}%
\providecommand \@@startlink[1]{}%
\providecommand \@@endlink[0]{}%
\providecommand \url  [0]{\begingroup\@sanitize@url \@url }%
\providecommand \@url [1]{\endgroup\@href {#1}{\urlprefix }}%
\providecommand \urlprefix  [0]{URL }%
\providecommand \Eprint [0]{\href }%
\providecommand \doibase [0]{http://dx.doi.org/}%
\providecommand \selectlanguage [0]{\@gobble}%
\providecommand \bibinfo  [0]{\@secondoftwo}%
\providecommand \bibfield  [0]{\@secondoftwo}%
\providecommand \translation [1]{[#1]}%
\providecommand \BibitemOpen [0]{}%
\providecommand \bibitemStop [0]{}%
\providecommand \bibitemNoStop [0]{.\EOS\space}%
\providecommand \EOS [0]{\spacefactor3000\relax}%
\providecommand \BibitemShut  [1]{\csname bibitem#1\endcsname}%
\let\auto@bib@innerbib\@empty
%</preamble>
\bibitem [{\citenamefont {Satir}\ and\ \citenamefont
  {Christensen}(2007)}]{Christensen2007}%
  \BibitemOpen
  \bibfield  {author} {\bibinfo {author} {\bibfnamefont {P.}~\bibnamefont
  {Satir}}\ and\ \bibinfo {author} {\bibfnamefont {S.~T.}\ \bibnamefont
  {Christensen}},\ }\href@noop {} {\bibfield  {journal} {\bibinfo  {journal}
  {Annual Review of Physiology}\ }\textbf {\bibinfo {volume} {69}},\ \bibinfo
  {pages} {377} (\bibinfo {year} {2007})}\BibitemShut {NoStop}%
\bibitem [{\citenamefont {Fulford}\ and\ \citenamefont
  {Blake}(1986)}]{fulford1986}%
  \BibitemOpen
  \bibfield  {author} {\bibinfo {author} {\bibfnamefont {G.~R.}\ \bibnamefont
  {Fulford}}\ and\ \bibinfo {author} {\bibfnamefont {J.~R.}\ \bibnamefont
  {Blake}},\ }\href@noop {} {\bibfield  {journal} {\bibinfo  {journal} {Journal
  of Theoretical Biology}\ }\textbf {\bibinfo {volume} {121}},\ \bibinfo
  {pages} {381} (\bibinfo {year} {1986})}\BibitemShut {NoStop}%
\bibitem [{\citenamefont {O'Callaghan}\ \emph {et~al.}(1999)\citenamefont
  {O'Callaghan}, \citenamefont {Sikand},\ and\ \citenamefont
  {Rutman}}]{ocallaghan_respiratory_1999}%
  \BibitemOpen
  \bibfield  {author} {\bibinfo {author} {\bibfnamefont {C.}~\bibnamefont
  {O'Callaghan}}, \bibinfo {author} {\bibfnamefont {K.}~\bibnamefont {Sikand}},
  \ and\ \bibinfo {author} {\bibfnamefont {A.}~\bibnamefont {Rutman}},\
  }\href@noop {} {\bibfield  {journal} {\bibinfo  {journal} {Pediatric
  research}\ }\textbf {\bibinfo {volume} {46}},\ \bibinfo {pages} {704}
  (\bibinfo {year} {1999})}\BibitemShut {NoStop}%
\bibitem [{\citenamefont {Randell}\ and\ \citenamefont
  {Boucher}(2006)}]{randell_effective_2006}%
  \BibitemOpen
  \bibfield  {author} {\bibinfo {author} {\bibfnamefont {S.~H.}\ \bibnamefont
  {Randell}}\ and\ \bibinfo {author} {\bibfnamefont {R.~C.}\ \bibnamefont
  {Boucher}},\ }\href@noop {} {\bibfield  {journal} {\bibinfo  {journal}
  {American journal of respiratory cell and molecular biology}\ }\textbf
  {\bibinfo {volume} {35}},\ \bibinfo {pages} {20} (\bibinfo {year}
  {2006})}\BibitemShut {NoStop}%
\bibitem [{\citenamefont {Del~Bigio}(1995)}]{DelBigio1995}%
  \BibitemOpen
  \bibfield  {author} {\bibinfo {author} {\bibfnamefont {M.~R.}\ \bibnamefont
  {Del~Bigio}},\ }\href@noop {} {\bibfield  {journal} {\bibinfo  {journal}
  {Glia}\ }\textbf {\bibinfo {volume} {14}},\ \bibinfo {pages} {1} (\bibinfo
  {year} {1995})}\BibitemShut {NoStop}%
\bibitem [{\citenamefont {Mirzadeh}\ \emph {et~al.}(2010)\citenamefont
  {Mirzadeh}, \citenamefont {Han}, \citenamefont {Soriano-Navarro},
  \citenamefont {Garc{\'\i}a-Verdugo},\ and\ \citenamefont
  {Alvarez-Buylla}}]{Mirzadeh2010}%
  \BibitemOpen
  \bibfield  {author} {\bibinfo {author} {\bibfnamefont {Z.}~\bibnamefont
  {Mirzadeh}}, \bibinfo {author} {\bibfnamefont {Y.-G.}\ \bibnamefont {Han}},
  \bibinfo {author} {\bibfnamefont {M.}~\bibnamefont {Soriano-Navarro}},
  \bibinfo {author} {\bibfnamefont {J.~M.}\ \bibnamefont
  {Garc{\'\i}a-Verdugo}}, \ and\ \bibinfo {author} {\bibfnamefont
  {A.}~\bibnamefont {Alvarez-Buylla}},\ }\href@noop {} {\bibfield  {journal}
  {\bibinfo  {journal} {The Journal of Neuroscience}\ }\textbf {\bibinfo
  {volume} {30}},\ \bibinfo {pages} {2600} (\bibinfo {year}
  {2010})}\BibitemShut {NoStop}%
\bibitem [{\citenamefont {Lyons}\ \emph {et~al.}(2006)\citenamefont {Lyons},
  \citenamefont {Saridogan},\ and\ \citenamefont {Djahanbakhch}}]{Lyons2006}%
  \BibitemOpen
  \bibfield  {author} {\bibinfo {author} {\bibfnamefont {R.}~\bibnamefont
  {Lyons}}, \bibinfo {author} {\bibfnamefont {E.}~\bibnamefont {Saridogan}}, \
  and\ \bibinfo {author} {\bibfnamefont {O.}~\bibnamefont {Djahanbakhch}},\
  }\href@noop {} {\bibfield  {journal} {\bibinfo  {journal} {Human reproduction
  update}\ }\textbf {\bibinfo {volume} {12}},\ \bibinfo {pages} {363} (\bibinfo
  {year} {2006})}\BibitemShut {NoStop}%
\bibitem [{\citenamefont {Wong}\ \emph {et~al.}(1993)\citenamefont {Wong},
  \citenamefont {Miller},\ and\ \citenamefont {Yeates}}]{wong_nature_1993}%
  \BibitemOpen
  \bibfield  {author} {\bibinfo {author} {\bibfnamefont {L.}~\bibnamefont
  {Wong}}, \bibinfo {author} {\bibfnamefont {I.~F.}\ \bibnamefont {Miller}}, \
  and\ \bibinfo {author} {\bibfnamefont {D.~B.}\ \bibnamefont {Yeates}},\
  }\href@noop {} {\bibfield  {journal} {\bibinfo  {journal} {Journal of Applied
  Physiology}\ }\textbf {\bibinfo {volume} {75}},\ \bibinfo {pages} {458}
  (\bibinfo {year} {1993})}\BibitemShut {NoStop}%
\bibitem [{\citenamefont {Wanner}\ \emph {et~al.}(1996)\citenamefont {Wanner},
  \citenamefont {Salath{\'e}},\ and\ \citenamefont
  {O'Riordan}}]{wanner_mucociliary_1996}%
  \BibitemOpen
  \bibfield  {author} {\bibinfo {author} {\bibfnamefont {A.}~\bibnamefont
  {Wanner}}, \bibinfo {author} {\bibfnamefont {M.}~\bibnamefont {Salath{\'e}}},
  \ and\ \bibinfo {author} {\bibfnamefont {T.~G.}\ \bibnamefont {O'Riordan}},\
  }\href@noop {} {\bibfield  {journal} {\bibinfo  {journal} {American journal
  of respiratory and critical care medicine}\ }\textbf {\bibinfo {volume}
  {154}},\ \bibinfo {pages} {1868} (\bibinfo {year} {1996})}\BibitemShut
  {NoStop}%
\bibitem [{\citenamefont {Davenport}\ and\ \citenamefont
  {Yoder}(2005)}]{davenport_incredible_2005}%
  \BibitemOpen
  \bibfield  {author} {\bibinfo {author} {\bibfnamefont {J.~R.}\ \bibnamefont
  {Davenport}}\ and\ \bibinfo {author} {\bibfnamefont {B.~K.}\ \bibnamefont
  {Yoder}},\ }\href@noop {} {\bibfield  {journal} {\bibinfo  {journal}
  {American Journal of Physiology-Renal Physiology}\ }\textbf {\bibinfo
  {volume} {289}},\ \bibinfo {pages} {F1159} (\bibinfo {year}
  {2005})}\BibitemShut {NoStop}%
\bibitem [{\citenamefont {Chopra}\ \emph {et~al.}(1977)\citenamefont {Chopra},
  \citenamefont {Taplin}, \citenamefont {Simmons},\ and\ \citenamefont
  {Elam}}]{chopra_measurement_1977}%
  \BibitemOpen
  \bibfield  {author} {\bibinfo {author} {\bibfnamefont {S.~K.}\ \bibnamefont
  {Chopra}}, \bibinfo {author} {\bibfnamefont {G.~V.}\ \bibnamefont {Taplin}},
  \bibinfo {author} {\bibfnamefont {D.~H.}\ \bibnamefont {Simmons}}, \ and\
  \bibinfo {author} {\bibfnamefont {D.}~\bibnamefont {Elam}},\ }\href@noop {}
  {\bibfield  {journal} {\bibinfo  {journal} {CHEST Journal}\ }\textbf
  {\bibinfo {volume} {71}},\ \bibinfo {pages} {155} (\bibinfo {year}
  {1977})}\BibitemShut {NoStop}%
\bibitem [{\citenamefont {Smith}\ \emph
  {et~al.}(2008{\natexlab{a}})\citenamefont {Smith}, \citenamefont {Gaffney},\
  and\ \citenamefont {Blake}}]{smith_modelling_2008}%
  \BibitemOpen
  \bibfield  {author} {\bibinfo {author} {\bibfnamefont {D.}~\bibnamefont
  {Smith}}, \bibinfo {author} {\bibfnamefont {E.}~\bibnamefont {Gaffney}}, \
  and\ \bibinfo {author} {\bibfnamefont {J.}~\bibnamefont {Blake}},\
  }\href@noop {} {\bibfield  {journal} {\bibinfo  {journal} {Respiratory
  physiology \& neurobiology}\ }\textbf {\bibinfo {volume} {163}},\ \bibinfo
  {pages} {178} (\bibinfo {year} {2008}{\natexlab{a}})}\BibitemShut {NoStop}%
\bibitem [{\citenamefont {Smith}\ \emph {et~al.}(2009)\citenamefont {Smith},
  \citenamefont {Gaffney},\ and\ \citenamefont
  {Blake}}]{smith_mathematical_2009}%
  \BibitemOpen
  \bibfield  {author} {\bibinfo {author} {\bibfnamefont {D.}~\bibnamefont
  {Smith}}, \bibinfo {author} {\bibfnamefont {E.}~\bibnamefont {Gaffney}}, \
  and\ \bibinfo {author} {\bibfnamefont {J.}~\bibnamefont {Blake}},\ }in\
  \href@noop {} {\emph {\bibinfo {booktitle} {Proceedings of the Royal Society
  of London A: Mathematical, Physical and Engineering Sciences}}}\ (\bibinfo
  {organization} {The Royal Society},\ \bibinfo {year} {2009})\ p.\ \bibinfo
  {pages} {rspa20090018}\BibitemShut {NoStop}%
\bibitem [{\citenamefont {Li}\ \emph {et~al.}(2012)\citenamefont {Li},
  \citenamefont {Chen}, \citenamefont {Ma}, \citenamefont {Tuo}, \citenamefont
  {Luo}, \citenamefont {Zhang}, \citenamefont {Sai}, \citenamefont {Liu},
  \citenamefont {Shen}, \citenamefont {Liu} \emph {et~al.}}]{li_methods_2012}%
  \BibitemOpen
  \bibfield  {author} {\bibinfo {author} {\bibfnamefont {W.-E.}\ \bibnamefont
  {Li}}, \bibinfo {author} {\bibfnamefont {W.}~\bibnamefont {Chen}}, \bibinfo
  {author} {\bibfnamefont {Y.-F.}\ \bibnamefont {Ma}}, \bibinfo {author}
  {\bibfnamefont {Q.-R.}\ \bibnamefont {Tuo}}, \bibinfo {author} {\bibfnamefont
  {X.-J.}\ \bibnamefont {Luo}}, \bibinfo {author} {\bibfnamefont
  {T.}~\bibnamefont {Zhang}}, \bibinfo {author} {\bibfnamefont {W.-B.}\
  \bibnamefont {Sai}}, \bibinfo {author} {\bibfnamefont {J.}~\bibnamefont
  {Liu}}, \bibinfo {author} {\bibfnamefont {J.}~\bibnamefont {Shen}}, \bibinfo
  {author} {\bibfnamefont {Z.-G.}\ \bibnamefont {Liu}},  \emph {et~al.},\
  }\href@noop {} {\bibfield  {journal} {\bibinfo  {journal} {Pfl{\"u}gers
  Archiv-European Journal of Physiology}\ }\textbf {\bibinfo {volume} {464}},\
  \bibinfo {pages} {671} (\bibinfo {year} {2012})}\BibitemShut {NoStop}%
\bibitem [{\citenamefont {Ding}\ \emph {et~al.}(2014)\citenamefont {Ding},
  \citenamefont {Nawroth}, \citenamefont {McFall-Ngai},\ and\ \citenamefont
  {Kanso}}]{ding2014}%
  \BibitemOpen
  \bibfield  {author} {\bibinfo {author} {\bibfnamefont {Y.}~\bibnamefont
  {Ding}}, \bibinfo {author} {\bibfnamefont {J.~C.}\ \bibnamefont {Nawroth}},
  \bibinfo {author} {\bibfnamefont {M.~J.}\ \bibnamefont {McFall-Ngai}}, \ and\
  \bibinfo {author} {\bibfnamefont {E.}~\bibnamefont {Kanso}},\ }\href@noop {}
  {\bibfield  {journal} {\bibinfo  {journal} {Journal of Fluid Mechanics}\
  }\textbf {\bibinfo {volume} {743}},\ \bibinfo {pages} {124} (\bibinfo {year}
  {2014})}\BibitemShut {NoStop}%
\bibitem [{\citenamefont {Fliegauf}\ \emph {et~al.}(2007)\citenamefont
  {Fliegauf}, \citenamefont {Benzing},\ and\ \citenamefont
  {Omran}}]{fliegauf2007}%
  \BibitemOpen
  \bibfield  {author} {\bibinfo {author} {\bibfnamefont {M.}~\bibnamefont
  {Fliegauf}}, \bibinfo {author} {\bibfnamefont {T.}~\bibnamefont {Benzing}}, \
  and\ \bibinfo {author} {\bibfnamefont {H.}~\bibnamefont {Omran}},\
  }\href@noop {} {\bibfield  {journal} {\bibinfo  {journal} {Nature Reviews
  Molecular Cell Biology}\ }\textbf {\bibinfo {volume} {8}},\ \bibinfo {pages}
  {880} (\bibinfo {year} {2007})}\BibitemShut {NoStop}%
\bibitem [{\citenamefont {Afzelius}(2004)}]{afzelius2004cilia}%
  \BibitemOpen
  \bibfield  {author} {\bibinfo {author} {\bibfnamefont {B.}~\bibnamefont
  {Afzelius}},\ }\href@noop {} {\bibfield  {journal} {\bibinfo  {journal} {The
  Journal of pathology}\ }\textbf {\bibinfo {volume} {204}},\ \bibinfo {pages}
  {470} (\bibinfo {year} {2004})}\BibitemShut {NoStop}%
\bibitem [{\citenamefont {Eloy}\ and\ \citenamefont {Lauga}(2012)}]{eloy2012}%
  \BibitemOpen
  \bibfield  {author} {\bibinfo {author} {\bibfnamefont {C.}~\bibnamefont
  {Eloy}}\ and\ \bibinfo {author} {\bibfnamefont {E.}~\bibnamefont {Lauga}},\
  }\href@noop {} {\bibfield  {journal} {\bibinfo  {journal} {Physical Review
  Letters}\ }\textbf {\bibinfo {volume} {109}},\ \bibinfo {pages} {038101}
  (\bibinfo {year} {2012})}\BibitemShut {NoStop}%
\bibitem [{\citenamefont {Osterman}\ and\ \citenamefont
  {Vilfan}(2011)}]{osterman2011}%
  \BibitemOpen
  \bibfield  {author} {\bibinfo {author} {\bibfnamefont {N.}~\bibnamefont
  {Osterman}}\ and\ \bibinfo {author} {\bibfnamefont {A.}~\bibnamefont
  {Vilfan}},\ }\href@noop {} {\bibfield  {journal} {\bibinfo  {journal}
  {Proceedings of the National Academy of Sciences}\ }\textbf {\bibinfo
  {volume} {108}},\ \bibinfo {pages} {15727} (\bibinfo {year}
  {2011})}\BibitemShut {NoStop}%
\bibitem [{\citenamefont {Guo}\ \emph {et~al.}(2014)\citenamefont {Guo},
  \citenamefont {Nawroth}, \citenamefont {Ding},\ and\ \citenamefont
  {Kanso}}]{guo2014}%
  \BibitemOpen
  \bibfield  {author} {\bibinfo {author} {\bibfnamefont {H.}~\bibnamefont
  {Guo}}, \bibinfo {author} {\bibfnamefont {J.}~\bibnamefont {Nawroth}},
  \bibinfo {author} {\bibfnamefont {Y.}~\bibnamefont {Ding}}, \ and\ \bibinfo
  {author} {\bibfnamefont {E.}~\bibnamefont {Kanso}},\ }\href@noop {}
  {\bibfield  {journal} {\bibinfo  {journal} {Physics of Fluids
  (1994-present)}\ }\textbf {\bibinfo {volume} {26}},\ \bibinfo {pages}
  {091901} (\bibinfo {year} {2014})}\BibitemShut {NoStop}%
\bibitem [{\citenamefont {Tucker}(1970)}]{tucker1970energetic}%
  \BibitemOpen
  \bibfield  {author} {\bibinfo {author} {\bibfnamefont {V.~A.}\ \bibnamefont
  {Tucker}},\ }\href@noop {} {\bibfield  {journal} {\bibinfo  {journal}
  {Comparative Biochemistry and Physiology}\ }\textbf {\bibinfo {volume}
  {34}},\ \bibinfo {pages} {841} (\bibinfo {year} {1970})}\BibitemShut
  {NoStop}%
\bibitem [{\citenamefont {Katsu-Kimura}\ \emph {et~al.}(2009)\citenamefont
  {Katsu-Kimura}, \citenamefont {Nakaya}, \citenamefont {Baba},\ and\
  \citenamefont {Mogami}}]{katsu2009}%
  \BibitemOpen
  \bibfield  {author} {\bibinfo {author} {\bibfnamefont {Y.}~\bibnamefont
  {Katsu-Kimura}}, \bibinfo {author} {\bibfnamefont {F.}~\bibnamefont
  {Nakaya}}, \bibinfo {author} {\bibfnamefont {S.~A.}\ \bibnamefont {Baba}}, \
  and\ \bibinfo {author} {\bibfnamefont {Y.}~\bibnamefont {Mogami}},\
  }\href@noop {} {\bibfield  {journal} {\bibinfo  {journal} {Journal of
  Experimental Biology}\ }\textbf {\bibinfo {volume} {212}},\ \bibinfo {pages}
  {1819} (\bibinfo {year} {2009})}\BibitemShut {NoStop}%
\bibitem [{\citenamefont {Sleigh}(1968)}]{sleigh1968}%
  \BibitemOpen
  \bibfield  {author} {\bibinfo {author} {\bibfnamefont {M.~A.}\ \bibnamefont
  {Sleigh}},\ }\href@noop {} {\bibfield  {journal} {\bibinfo  {journal}
  {Symposia of the Society for Experimental Biology}\ }\textbf {\bibinfo
  {volume} {22}},\ \bibinfo {pages} {131} (\bibinfo {year} {1968})}\BibitemShut
  {NoStop}%
\bibitem [{\citenamefont {Taylor}(1951)}]{taylor1951}%
  \BibitemOpen
  \bibfield  {author} {\bibinfo {author} {\bibfnamefont {G.~I.}\ \bibnamefont
  {Taylor}},\ }\href@noop {} {\bibfield  {journal} {\bibinfo  {journal}
  {Proceedings of the Royal Society of London. Series A}\ }\textbf {\bibinfo
  {volume} {209}},\ \bibinfo {pages} {447} (\bibinfo {year}
  {1951})}\BibitemShut {NoStop}%
\bibitem [{\citenamefont {Eshel}\ and\ \citenamefont
  {Brokaw}(1987)}]{eshel1987new}%
  \BibitemOpen
  \bibfield  {author} {\bibinfo {author} {\bibfnamefont {D.}~\bibnamefont
  {Eshel}}\ and\ \bibinfo {author} {\bibfnamefont {C.~J.}\ \bibnamefont
  {Brokaw}},\ }\href@noop {} {\bibfield  {journal} {\bibinfo  {journal} {Cell
  motility and the cytoskeleton}\ }\textbf {\bibinfo {volume} {7}},\ \bibinfo
  {pages} {160} (\bibinfo {year} {1987})}\BibitemShut {NoStop}%
\bibitem [{\citenamefont {Bayly}\ \emph {et~al.}(2010)\citenamefont {Bayly},
  \citenamefont {Lewis}, \citenamefont {Kemp}, \citenamefont {Pless},\ and\
  \citenamefont {Dutcher}}]{bayly2010efficient}%
  \BibitemOpen
  \bibfield  {author} {\bibinfo {author} {\bibfnamefont {P.}~\bibnamefont
  {Bayly}}, \bibinfo {author} {\bibfnamefont {B.}~\bibnamefont {Lewis}},
  \bibinfo {author} {\bibfnamefont {P.}~\bibnamefont {Kemp}}, \bibinfo {author}
  {\bibfnamefont {R.}~\bibnamefont {Pless}}, \ and\ \bibinfo {author}
  {\bibfnamefont {S.}~\bibnamefont {Dutcher}},\ }\href@noop {} {\bibfield
  {journal} {\bibinfo  {journal} {Cytoskeleton}\ }\textbf {\bibinfo {volume}
  {67}},\ \bibinfo {pages} {56} (\bibinfo {year} {2010})}\BibitemShut {NoStop}%
\bibitem [{\citenamefont {Sartori}\ \emph {et~al.}(2015)\citenamefont
  {Sartori}, \citenamefont {Geyer}, \citenamefont {Scholich}, \citenamefont
  {J{\"u}licher},\ and\ \citenamefont {Howard}}]{sartori2015dynamic}%
  \BibitemOpen
  \bibfield  {author} {\bibinfo {author} {\bibfnamefont {P.}~\bibnamefont
  {Sartori}}, \bibinfo {author} {\bibfnamefont {V.}~\bibnamefont {Geyer}},
  \bibinfo {author} {\bibfnamefont {A.}~\bibnamefont {Scholich}}, \bibinfo
  {author} {\bibfnamefont {F.}~\bibnamefont {J{\"u}licher}}, \ and\ \bibinfo
  {author} {\bibfnamefont {J.}~\bibnamefont {Howard}},\ }\href@noop {}
  {\bibfield  {journal} {\bibinfo  {journal} {arXiv preprint arXiv:1511.04270}\
  } (\bibinfo {year} {2015})}\BibitemShut {NoStop}%
\bibitem [{\citenamefont {Cortez}\ \emph {et~al.}(2005)\citenamefont {Cortez},
  \citenamefont {Fauci},\ and\ \citenamefont {Medovikov}}]{cortez2005}%
  \BibitemOpen
  \bibfield  {author} {\bibinfo {author} {\bibfnamefont {R.}~\bibnamefont
  {Cortez}}, \bibinfo {author} {\bibfnamefont {L.}~\bibnamefont {Fauci}}, \
  and\ \bibinfo {author} {\bibfnamefont {A.}~\bibnamefont {Medovikov}},\
  }\href@noop {} {\bibfield  {journal} {\bibinfo  {journal} {Physics of Fluids
  (1994-present)}\ }\textbf {\bibinfo {volume} {17}},\ \bibinfo {pages}
  {031504} (\bibinfo {year} {2005})}\BibitemShut {NoStop}%
\bibitem [{\citenamefont {Ainley}\ \emph {et~al.}(2008)\citenamefont {Ainley},
  \citenamefont {Durkin}, \citenamefont {Embid}, \citenamefont {Boindala},\
  and\ \citenamefont {Cortez}}]{ainley2008}%
  \BibitemOpen
  \bibfield  {author} {\bibinfo {author} {\bibfnamefont {J.}~\bibnamefont
  {Ainley}}, \bibinfo {author} {\bibfnamefont {S.}~\bibnamefont {Durkin}},
  \bibinfo {author} {\bibfnamefont {R.}~\bibnamefont {Embid}}, \bibinfo
  {author} {\bibfnamefont {P.}~\bibnamefont {Boindala}}, \ and\ \bibinfo
  {author} {\bibfnamefont {R.}~\bibnamefont {Cortez}},\ }\href@noop {}
  {\bibfield  {journal} {\bibinfo  {journal} {Journal of Computational
  Physics}\ }\textbf {\bibinfo {volume} {227}},\ \bibinfo {pages} {4600}
  (\bibinfo {year} {2008})}\BibitemShut {NoStop}%
\bibitem [{\citenamefont {Liron}(1978)}]{liron1978fluid}%
  \BibitemOpen
  \bibfield  {author} {\bibinfo {author} {\bibfnamefont {N.}~\bibnamefont
  {Liron}},\ }\href@noop {} {\bibfield  {journal} {\bibinfo  {journal} {Journal
  of Fluid Mechanics}\ }\textbf {\bibinfo {volume} {86}},\ \bibinfo {pages}
  {705} (\bibinfo {year} {1978})}\BibitemShut {NoStop}%
\bibitem [{\citenamefont {Smith}\ \emph
  {et~al.}(2008{\natexlab{b}})\citenamefont {Smith}, \citenamefont {Blake},\
  and\ \citenamefont {Gaffney}}]{smith2008}%
  \BibitemOpen
  \bibfield  {author} {\bibinfo {author} {\bibfnamefont {D.~J.}\ \bibnamefont
  {Smith}}, \bibinfo {author} {\bibfnamefont {J.~R.}\ \bibnamefont {Blake}}, \
  and\ \bibinfo {author} {\bibfnamefont {E.~A.}\ \bibnamefont {Gaffney}},\
  }\href@noop {} {\bibfield  {journal} {\bibinfo  {journal} {Journal of The
  Royal Society Interface}\ }\textbf {\bibinfo {volume} {5}},\ \bibinfo {pages}
  {567} (\bibinfo {year} {2008}{\natexlab{b}})}\BibitemShut {NoStop}%
\bibitem [{\citenamefont {Machin}(1958)}]{machin1958wave}%
  \BibitemOpen
  \bibfield  {author} {\bibinfo {author} {\bibfnamefont {K.}~\bibnamefont
  {Machin}},\ }\href@noop {} {\bibfield  {journal} {\bibinfo  {journal} {J.
  exp. Biol}\ }\textbf {\bibinfo {volume} {35}},\ \bibinfo {pages} {796}
  (\bibinfo {year} {1958})}\BibitemShut {NoStop}%
\bibitem [{\citenamefont {Brennen}\ and\ \citenamefont
  {Winet}(1977)}]{brennen1977}%
  \BibitemOpen
  \bibfield  {author} {\bibinfo {author} {\bibfnamefont {C.}~\bibnamefont
  {Brennen}}\ and\ \bibinfo {author} {\bibfnamefont {H.}~\bibnamefont
  {Winet}},\ }\href@noop {} {\bibfield  {journal} {\bibinfo  {journal} {Annual
  Review of Fluid Mechanics}\ }\textbf {\bibinfo {volume} {9}},\ \bibinfo
  {pages} {339} (\bibinfo {year} {1977})}\BibitemShut {NoStop}%
\bibitem [{\citenamefont {Hines}\ and\ \citenamefont {Blum}(1983)}]{hines1983}%
  \BibitemOpen
  \bibfield  {author} {\bibinfo {author} {\bibfnamefont {M.}~\bibnamefont
  {Hines}}\ and\ \bibinfo {author} {\bibfnamefont {J.~J.}\ \bibnamefont
  {Blum}},\ }\href@noop {} {\bibfield  {journal} {\bibinfo  {journal}
  {Biophysical Journal}\ }\textbf {\bibinfo {volume} {41}},\ \bibinfo {pages}
  {67} (\bibinfo {year} {1983})}\BibitemShut {NoStop}%
\bibitem [{\citenamefont {Gittes}\ \emph {et~al.}(1993)\citenamefont {Gittes},
  \citenamefont {Mickey}, \citenamefont {Nettleton},\ and\ \citenamefont
  {Howard}}]{gittes1993flexural}%
  \BibitemOpen
  \bibfield  {author} {\bibinfo {author} {\bibfnamefont {F.}~\bibnamefont
  {Gittes}}, \bibinfo {author} {\bibfnamefont {B.}~\bibnamefont {Mickey}},
  \bibinfo {author} {\bibfnamefont {J.}~\bibnamefont {Nettleton}}, \ and\
  \bibinfo {author} {\bibfnamefont {J.}~\bibnamefont {Howard}},\ }\href@noop {}
  {\bibfield  {journal} {\bibinfo  {journal} {The Journal of cell biology}\
  }\textbf {\bibinfo {volume} {120}},\ \bibinfo {pages} {923} (\bibinfo {year}
  {1993})}\BibitemShut {NoStop}%
\bibitem [{\citenamefont {Lauga}\ and\ \citenamefont
  {Powers}(2009)}]{lauga2009}%
  \BibitemOpen
  \bibfield  {author} {\bibinfo {author} {\bibfnamefont {E.}~\bibnamefont
  {Lauga}}\ and\ \bibinfo {author} {\bibfnamefont {T.~R.}\ \bibnamefont
  {Powers}},\ }\href@noop {} {\bibfield  {journal} {\bibinfo  {journal}
  {Reports on Progress in Physics}\ }\textbf {\bibinfo {volume} {72}},\
  \bibinfo {pages} {096601} (\bibinfo {year} {2009})}\BibitemShut {NoStop}%
\end{thebibliography}%

\end{document}